\documentclass[twoside,12pt,a4paper]{article}
\usepackage[english]{babel}
\usepackage[utf8]{inputenc}
\usepackage[twoside,margin=1.2in]{geometry}
\usepackage{graphicx}
\usepackage{amsmath}
\usepackage[square,sort]{natbib}
\usepackage{framed, color}
\definecolor{shadecolor}{rgb}{0.9, 0.9, 0.9}
\newenvironment{definition}[1][Definition]{\begin{trivlist}
	\item[\hskip \labelsep {\bfseries #1}]}{\end{trivlist}}

\newcommand{\bu}[1]{\mbox{$\mathbf{#1}$}}
\usepackage{xspace}
\newcommand{\R}{\textnormal{\sffamily\bfseries R}\xspace}

\usepackage{Sweave}
\begin{document}
\setkeys{Gin}{width=0.8\textwidth}

\begin{center}
  {\Large Parameter space exploration of ecological models}

\begin{table}[hbt]{\small
\begin{tabular}{ll}
Authors        & Andre Chalom \\
               & Paulo Inacio Prado \\
Date           & 06/30/2015 \\
Affiliation    & Department of Ecology, \\
	       & University of Sao Paulo 
\end{tabular} }
\end{table}
\end{center}

\mbox{}\vspace{-14mm}\mbox{}

\begin{abstract}
There is a growing trend in the use of mathematical modeling tools in the
study of many areas of the biological sciences. 

The use of computer models in science is increasing, specially in fields where laboratory
experiments are too complex or too costly, like ecology.

Questions of efficient, systematic and error-proof exploration of parameter spaces are
are of great importance to better understand, estimate confidences and make use of the 
output from these models. 

We present a survey of the proposed methods to answer these questions, with emphasis
on the Latin Hypercube Sampling and focusing on quantitative analysis of the results. We also
compare analytical results for sensitivity and uncertainty, where relevant, to LHS results.

This document contains a revision about the state-of-art uncertainty and sensitivity analyses,
with a practical example of applying the described techniques to two models of structured population growth.

During the progress of this work, a package of \R functions was developed
to facilitate the real world use of the above theoretical tools, freely available at
http://cran.r-project.org/web/packages/pse.

	  \par
	  \vspace{1em}
\end{abstract}

\newpage
\tableofcontents
\newpage

\section{Introduction}\label{Introduction}
There is a growing trend in the use of mathematical modeling tools in the
study of many areas of the biological sciences. The use of models is essential
as they present an opportunity to address questions that are impossible or
impractical to answer in either in purely theoretical analyses or in field 
or laboratory experiments, and to identify the most important processes which
should then be investigated by experiments. One compelling example is made by the
Individual Based
Models (IBM), which represent individuals that move and interact in space,
according to some decision-making rules. 
These models permit a great level of detail and realism to be included,
as well as linking multiple levels of complexity in a system.

On the other hand, more realistic models employ a vast selection of input
parameters, from temperature and rainfall to metabolic and encounter rates,
which may be difficult to accurately measure. Moreover, one may be interested
in estimating how much predictions for a model fitted in one place or to one species
may be extrapolated to different places or species.
While variations in some of those
parameters will have negligible impact on the model output, other parameters
may profoundly impact the validity of a model's predictions, and it may be
impossible to determine {\em a priori} which are the most important parameters.
A {\em na\"ive} approach would consist on the evaluation of the model at all possible 
combinations of parameters, however this would require a prohibitive number of
model runs, specially considering that a single run of those models may take
days to complete. Our challenge then consists in providing the best estimates
for the importance of the several parameters, requiring the least number of
model runs.

Some models are either expressed or can be reasonably approximated by analytical
functions. Such is the case for the matrix population models, for which a wide
range of analytical tools are available to examine the uncertainty and
sensitivity of parameters\citep{Caswell89}. In the general case, however, there
is no possible formulation of the model in a closed equation, so analytical 
methods are not possible.

The disciplines of uncertainty and sensitivity analysis have been developed in
the context of the physical sciences and engineering, and have been greatly
developed in the 1980 and 1990 decades \citep{McKay79, Saltelli04, Florian92,
Helton03, Helton05, Archer97, Huntington98, Ye98, Kleijnen99, Smith02, 
ImanConover82, ImanConover87, Morris91, Saltelli99}.

More recently, these analyses have been 
successfully applied to biological models, in order to explore the possible 
outcomes from the model output, estimate their
probability distribution and the dependency of the output 
on different combinations of parameters, and to assess which parameters 
require more experimental effort in order to be more confidently estimated. 
This kind of parameter space exploration is considered a fundamental step
prior to using the model in management decisions \citep{Bart95}. 

One approach to the parameter space exploration, which will be described here,
is to generate samples from the parameter space, run the model with these 
samples, and analyze the qualitative or quantitative differences in the
model output. In this context, the sampling of the parameter space may 
be regarded as a bridge between the modeling of the system and the inference
problem of acquiring information about the whole parameter space, having only
access to a subset of that information. This inference can be done in the light
of any of the statistical schools.

Section \ref{Sampling} will present the sampling techniques,
with emphasis on the Latin Hypercube method, while section 
\ref{QuantAnal} will present some tools for the quantitative analyses.
We should emphasize that the analyses tools are not coupled with the sampling
techniques used: one can, in principle, use the sampling techniques described
and other analyses tools, or apply the analyses described here to a more
general class of sampling methods.
We also present two examples of the sampling and analysis used in section \ref{Leslie}
and \ref{Tribolium}.
Then, we briefly review some relevant research papers which
have used such techniques in the exploration of ecological models in section
\ref{Studies}.

\subsection{Parameter spaces}\label{ps}
In order to better pose our questions, we need first to discuss some
properties of the parameter space (or PS for short) of our models.

The parameters (or inputs) are quantities $x_1, x_2, \cdots, x_m$ which
will be used to run the model. In our discussion, we will assume that
all the $x_i$ are real valued.
These quantities are unknown, and one first
challenge is to determine which set of values better
fit a model to the available data, which is the subject of linear and
nonlinear estimation.

However, the same model may be
parametrized in different ways, as discussed by \cite{Ross90}. For
example, in population ecology, 
the logistic growth equation may be represented with two
parameters $r$ and $K$ as:
\begin{equation}
		\frac{dN}{dt} = rN\left(1-\frac{N}{K}\right)
		\label{logrk}
\end{equation}
However, the same equation may be written as
\begin{equation}
		\frac{dN}{dt} = \alpha N + \beta N^2
		\label{logalphabeta}
\end{equation}
Here, the two parameters are $\alpha = r$ and $\beta = -r/K$. 
While the first equation is far more commonly used in the biological 
context, both are equivalent, and using one or the other model
is simply a matter of choice.

There are many other ways of writing this equation, and one of special
interest when trying to fit real data is in terms of orthogonal 
polynomials, such as:
\begin{equation}
		\frac{dN}{dt} = \theta_0 + \theta_1(x-\bar x) + \theta_2((x-\bar x)^2 - (\overline {x - \bar {x}})^2)
		\label{logtheta}
\end{equation}
Where $\theta_0$ can be calculated from $\theta_1$ and $\theta_2$. This
more complicated equation has several numerical advantages over the previous,
as the parameters $\theta_1$ and $\theta_2$ can be estimated with much more
accuracy, and will not be correlated (as is the case with $\alpha$ and
$\beta$, as well as $r$ and $K$). However, these parameters are hard to
interpret in biological
terms. 

These different equations illustrate the existence of {\em interpretable}
(Eq. \ref{logrk}), {\em defining}, or algebraic (Eq. \ref{logalphabeta})
and {\em computing} (Eq. \ref{logtheta}) parameters. Most of the times,
it would be preferable to estimate the values that best fit some data by
using computing parameters, and then to transform them to interpretable
parameters in order to present the results.

Also, it should be mentioned that the parameter space may be constrained.
This will have an impact on some of the available sampling 
and analysis techniques.
The simplest constraint is requiring some parameter to be positive or
negative. Also, there may be combinations of values that are meaningless.
For example, if we model a community with $N$ individuals and $S$ species,
the number of individuals and species, considered on their own, may be
any positive number. However, it is clear that the number of species may
not be bigger than the number of individuals, which imposes the
condition $S \leq N$. This condition is called a {\em constraint}, and
limits the values that the parameter vector
may assume.

If we consider the m-dimensional space consisting on all possible 
combination of values for the parameters, our parameter space will be the
subset of this space that respects all our constraints. For example, consider
that the parameters we are interested are two angles of a triangle. 
In this case, the sum of the angles must be less than 180 degrees, $a_1+a_2 <180 \circ$.
Clearly, this
parameter space is not square, in the sense that, if we define the ranges of
the variables $a_1$ and $a_2$ independently as $(0,180)$, not all combinations
of parameters will be meaningful. What can be done in this case is to create
a new parameter $\hat{a_1}$, defined as
\begin{equation}
	\hat{a_1} = \frac{a_1}{180-a_2}
	\label{hata1}
\end{equation}

\begin{figure}[htbp]
	\begin{center}
\includegraphics{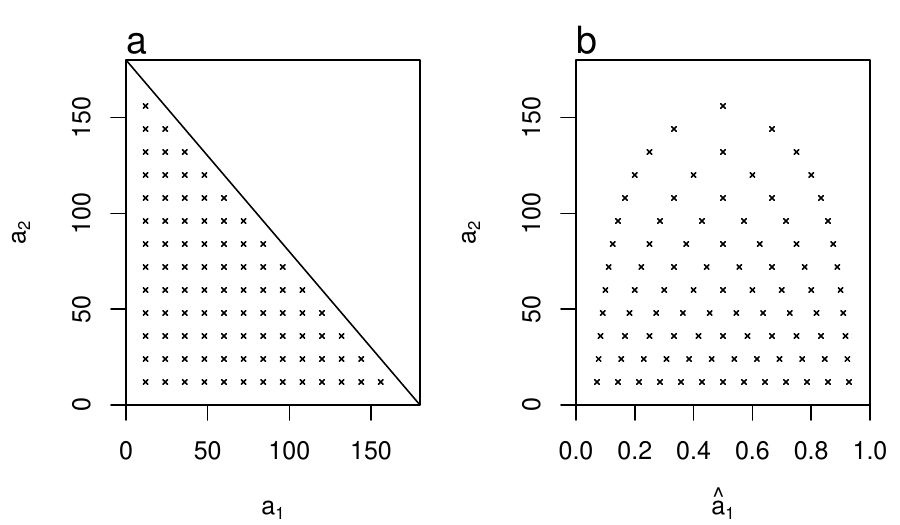}
	\end{center}
	\caption{a. The constrained parameter space considered, 
	with the line representing $a_1+a_2=180$. The symbols represent 
	a uniform sample taken from the space. b. The transformed parameter 
	space $\hat{a_1}, a_2$ (see Eq. \ref{hata1}), showing the same sampled
	points.}
	\label{fig:Trans}
\end{figure}

This new parameter varies between 0 and 1, and all combinations of 
$\hat{a_1}, a_2$ are points from our parameter space. Now, care must be 
exercised after applying such transformations in order to preserve the marginal
distributions from the original variables, as exemplified on figure 
\ref{fig:Trans}.

\setkeys{Gin}{width=1.0\textwidth}
\setkeys{Gin}{width=0.8\textwidth}

Another related concept that should not be confused with the constraints
is the correlation between variables. For example, acidic soils are likely 
to have a lower cation exchange capacity (CEC), and more alkaline soils are 
likely to have a 
larger CEC \citep{Sparks03}. Thus, those variables are {\em correlated}. 
Correlations have a profound impact on some analysis, however, they are
difficult
to measure, and data on correlations are not generally available on the 
literature. 
We 
will return to questions related to correlations in parameter
spaces in section \ref{ext}.

\subsection{Applications of parameter space exploration}\label{PSE}
Next, we turn our attention to the kind of problems we might want to address
with the exploration of the parameter space. First, the simplest case is asking
``is there a region of my parameter space where condition X holds?'' This 
condition might be, for example, the extinction or coexistence of species, 
some pattern of distribution or abundance of species. We also might be 
interested in mapping where are these regions. In complex models, where several 
different regions might exist where the qualitative results of the models are 
very different, we may ask how many of these regions are there, as well as map
the frontiers between them. For a detailed discussion of this approach, see
the PSExplorer software\citep{Tung10}.

Another class of problems arises when the model produces some quantitative 
response, and we are interested in determining the dependency of this response 
to the input parameters. For example, when modeling the dynamics of a 
population, we might want to know how the final population varies with each 
of the input parameters. In this context of quantitative analysis, the 
questions are divided in two classes: first, how much the variation of the 
input parameters is translated into the total variation of the results, 
which is the topic of uncertainty analysis, 
and second, how much of the variation in the results can be ascribed to the 
variation of each individual parameter, which is the topic of sensitivity 
analysis \citep{Helton03, Helton05}. We will present the techniques and results
from both uncertainty and sensitivity analysis in section \ref{QuantAnal}.

Also, the model which we wish to analyze can be any function of the input parameters.
In particular, there are three classes of models that can be used. First,
the model may be a complex mathematical function (for example, defined
by a differential equation). Second, the model may be a simulation model, like
an IBM. Third, the model may be the result of fitting a statistical model.

All these problems may be formulated in a general way, defining some response 
from the model $\bu{Y}$ as a function of the input parameter vector $\bu{x}$:

\begin{equation}
	\bu{Y} = \bu{f}(\bu{x})
	\label{genmodel}
\end{equation}

In the equation \ref{genmodel}, all the quantities are vectors, indicated by
the boldface. Here, $\bu{x} = [ x_1,x_2,\dots,x_m ]$ represent the parameters
to the model $\bu{f}$, and $\bu{Y} = [ y_1,y_2,\dots,y_n ]$ represent the some
quantitative responses from the model. In some sections, we will discuss the
response as a single value $y$, without loss of generality.

\begin{figure}[htpb]
	\begin{center}
	
\includegraphics{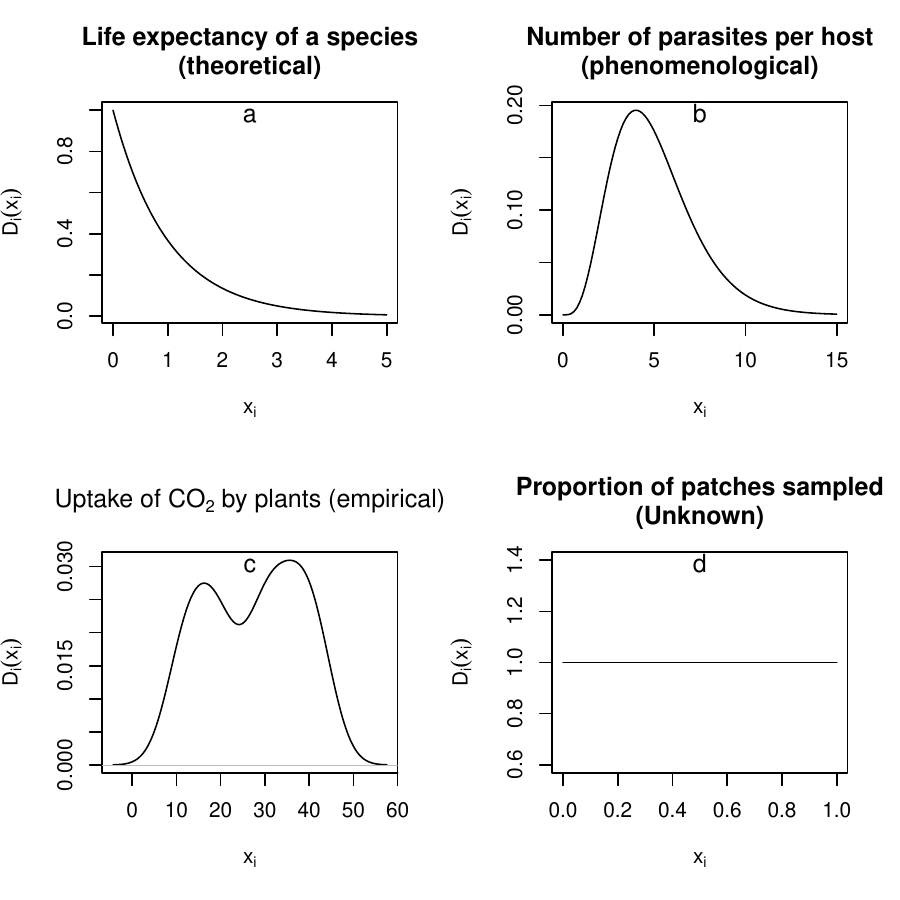}
	\end{center}
	\caption{Four different possibilities for choosing the distributions $D_i$. Panel ``a'' shows 
	the exponential distribution of life expectancy, which can be deduced
	from theory \citep{Cole54}. Panel ``b'' shows a gamma distribution,
	which can be used to model the number of parasites per host, but has
	no theoretical derivation \citep{Bolker08}. Panel ``c'' shows data 
	from an empirical study on the $CO_2$ uptake from plants 
	\citep{Potvin90}, and panel ``d'' shows an example where no prior
	information can be used.}
	\label{fig:Di}
\end{figure}

Each of the input parameters $x_i$ is associated with a probability distribution
$D_i(x)$, which represent our degree of knowledge about the values that $x_i$ may
assume (see figure \ref{fig:Di} for examples); \cite{Berger85} provides
a more detailed discussion).

Taken together, all the distributions $D_i$ form the {\em joint probability
distribution} of the parameters, $\bu{D}(\bu{x})$. This function takes into
account not only the individual distribution of each parameter, but also
all the correlation terms between them\footnote{For clarity, it should be
noted that the joint probability distribution is {\em not} used explicitly
in the methods discussed here. Only the marginal distributions and correlation
terms, if necessary, are explicitly used.}. 

In very simple models, it may be possible to analytically deduce the behavior
of the model response taken at each point of the joint distribution of parameters.
In the general case, however, this is impossible, and a way of investigating the
model is to choose some points from the joint distribution and analyzing the 
model at each point. Section \ref{Sampling} will present some strategies for
choosing these points.

\section{Sampling Techniques}\label{Sampling}
There are several strategies that can be used to choose the samples from the
parameter space that will be used as input to our model of interest. Here,
we will present some of them, along with their limitations, to justify our 
choice for the Latin Hypercube Sampling, which we will describe in section
\ref{LHS}.

One way of exploring the parameter space is by discretizing every distribution
and running the model for every
possible combination of values for all parameters. 
This is called full parameter space exploration, as done 
by \cite{Turchin97}, and although it 
possesses many advantages, it may become very costly in terms of computer time. 
In addition, the number of possible combinations increases exponentially with 
the number of parameter dimensions considered.

To circumvent the exponential increase in the number of samples, it is usual
to explore the parameter space in the following fashion: holding all but one
parameter constant, we analyze how the output of a model is affected by one
parameter dimension at a time (as done by \cite{Yang08}). This analysis is
referred to as individual parameter disturbance.
This kind of analysis
is, however, limited by the fact that the combinations of changed parameters
may give rise to complex and unexpected behaviors. 
Often, algorithms based on individual parameter disturbance are used as a first 
step in order to discriminate between parameters
that may have a substantial impact on the output, and parameters that are less
relevant (but see \citep{Morris91} for an alternative). 

Another viable option would be to
chose $N$ random samples from the entire space, in order to analyze both
the effect of each parameter and the combined effect of changing any
combined number of parameters. This sampling scheme is called random 
sampling, or Monte Carlo sampling,
and has been applied to many biological models \citep{Letcher96}.
One important feature of the Monte Carlo sampling is that its accuracy
does not depend on the number of dimensions of the problem
\citep{MacKay03}.

\begin{figure}[htbp]
	\begin{center}
\includegraphics{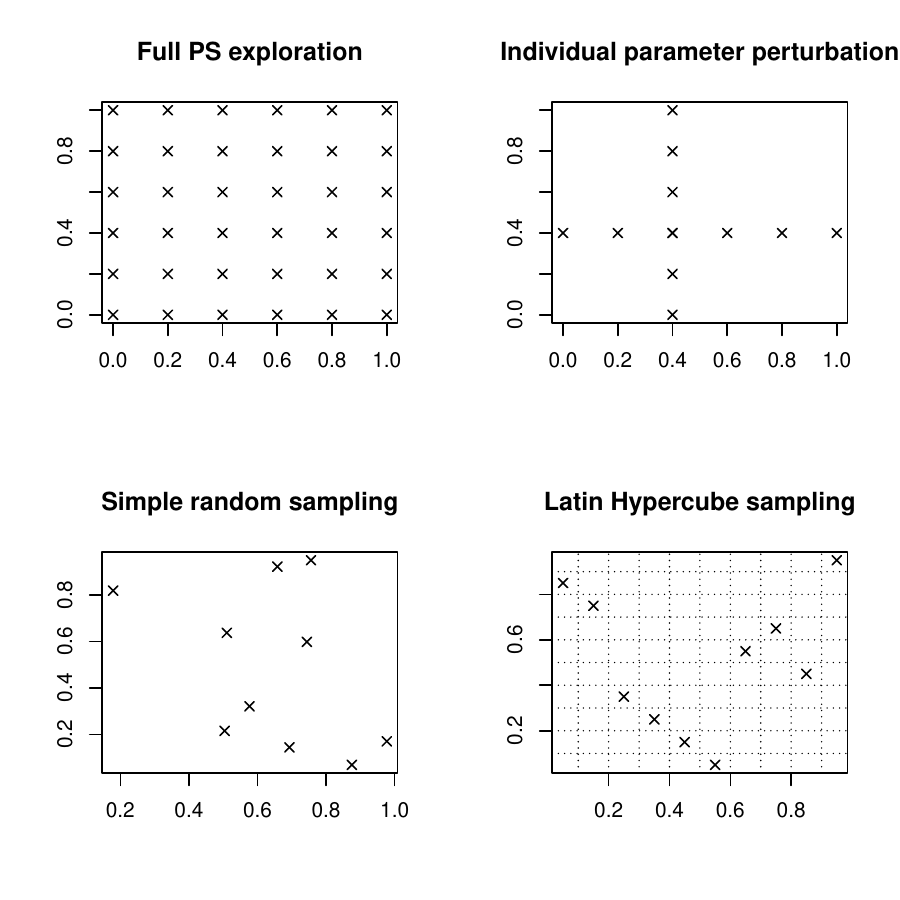}
	\end{center}
	\caption{Illustration of four sampling methods. While the
	full parameter space exploration is clearly representative
	of the whole space, it requires a very large number of samples.
	The individual parameter perturbation chooses samples by
	holding one parameter constant and varying the other, and
	clearly cannot take into account interactions between 
	parameters. The random sampling uses information about the
	whole parameter space with a small number of samples, but
	can oversample some regions while under sampling others. The
	Latin Hypercube (section \ref {LHS}) samples all the 
	intervals with equal intensity.}
	\label{fig:SamplingMethods}
\end{figure}

Stratified sampling strategies, which are a special case of Monte 
Carlo sampling, consist in strategies for choosing these random samples while,
at the same, making sure that each of the subdivisions (or {\em strata}) of
the distribution are well represented.
As shown by \cite{McKay79}, the estimates of statistical properties (such
as the mean or the variance) of the model output are better represented by 
stratified random sampling than by simple random sampling
(see figure \ref{fig:SamplingMethods} for examples). As we shall see in the
next session, the Latin Hypercube sampling is a practical and easy to understand
stratified sampling strategy.

Another class of Monte Carlo methods that should be mentioned here is the Markov
Chain Monte Carlo (MCMC), which is also used on similar analyses \citep{MacKay03}. This
methods consists in generating a sequence of points $\{\bu{x}^{(t)}\}$ from
the parameter space whose distribution {\em converges} to the joint
probability distribution $\bu{D}(\bu{x})$, and in which each sample $\bu{x}^{(t)}$
is chosen based on the previous $\bu{x}^{(t-1)}$. MCMC methods perform better
than LHS methods for estimating the distribution of the model responses,
however, they require a number of model runs which is orders of magnitude higher
than LHS requirements.

\subsection{Latin Hypercube: Definition and use}\label{LHS}
In this section, we describe the Latin Hypercube Sampling, and show how it can
be used to efficiently solve the questions posed in section \ref{Introduction}.
We also
discuss what are the available methods for obtaining the LHS. 

Firstly, let us define, in the context of statistical sampling, what is a 
Latin Square:
\begin{definition}
If we divide each side in a square in $N$ intervals, and then take samples
from the square, the resulting square will be called
Latin if and only if there is exactly one sample in each row and each column.
\end{definition}

A Latin Hypercube is simply the generalization of the Latin
Square to an arbitrary number of dimensions $m$. 
It should be noted, then, that the number of samples $N$ is fixed {\em a priori},
and does not depend on the number of parameters considered.

We will now construct the Latin Hypercube. Let's fix our attention in one parameter
dimension $i$ of the parameter space.
The first step we should take is to divide the range of $x_i$
in $N$ equally probable intervals. In order to do so, we will turn our attention
to the probability distribution of $x_i$, defined on section \ref{ps} as $D_i$.
Recall that this probability distribution must be chosen in a way that represents our
current understanding of the biology of the given system. This function might
be estimated by an expert in the field, it might represent a data set from
field or laboratory work, or in some cases it may be simply the broadest
possible set of parameters, in some cases where the actual values are unknown
or experiments are unfeasible (see fig. \ref{fig:Di}). 

In possession of the distribution function $D_i$, we must sample one point
from each equally probable interval. There are two approaches used here:
it is possible to choose a random value from within the interval 
\citep{McKay79}, or instead, we can use the midpoint from each interval 
\citep{Huntington98}. As the statistical properties of the generated samples
are very similar, we will use the second approach here.

The integral of the distribution function is called the cumulative distribution
function $F_i(x)$. This function relates the values $x$ that the parameter may assume
with the probability $p$ that the parameter is less than or equal to $x$.
We will refer to the inverse of the cumulative distribution function, $F_i^{-1}$,
as the quantile function of the parameter $x_i$, as it associates every probability
value $p$ in the range $(0,1)$ to the value $x$ such that $P(x_i \leq x) = p$.
We divide the range $(0,1)$ in $N$ intervals of size $1/N$, and use this quantile
function to determine the $x$ values as the midpoints of each interval. 
Summarizing, we take the $N$ points, represented as $x_{i,k}$, $k \in [1,N]$, 
from the inverse cumulative distribution $F_i^{-1}(x)$ as\footnote{This formula is
given for simplicity; see \citep{Huntington98} for an alternative with better numerical
properties}:

\begin{equation}
	x_{i,k} = F_i^{-1}\left( \frac{k-0.5}{N}\right)
	\label{inverseCDF}
\end{equation}

\begin{figure}[htpb]
	\begin{center}
\includegraphics{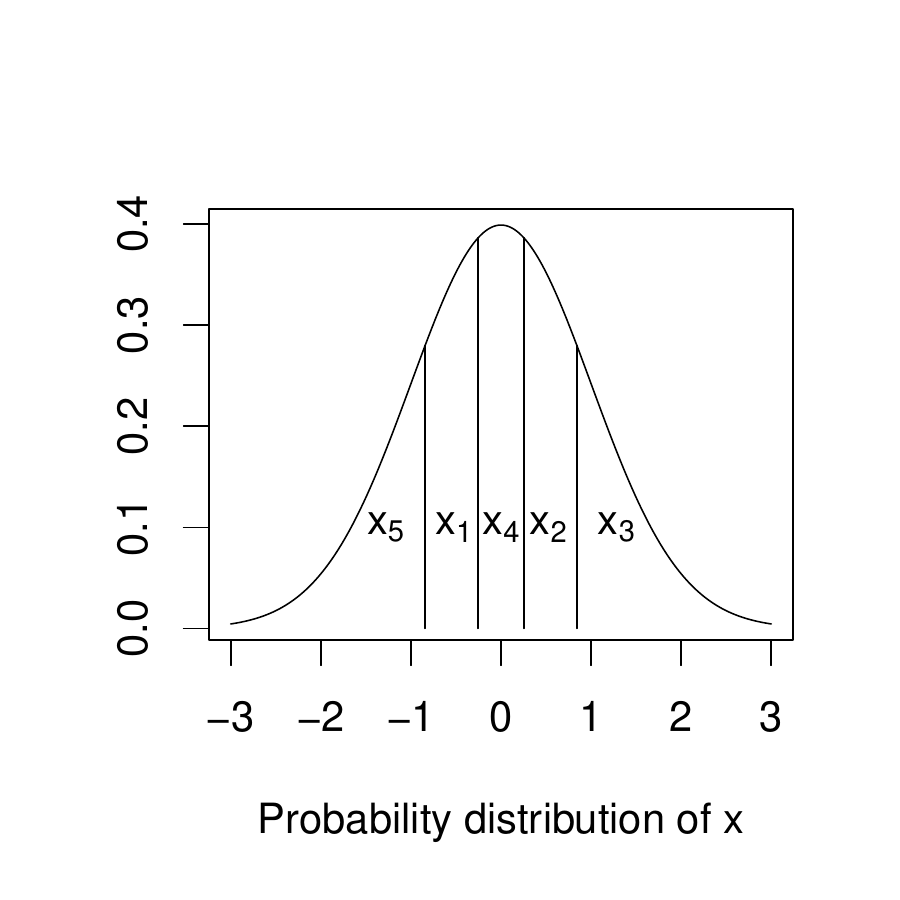}
	\end{center}
	\caption{Sample normal probability distribution, with 5 samples 
	collected from regions with same probability and shuffled. 
	Note that the first sample correspond to the 
	second interval, the second sample correspond to the fourth interval,
	and so on.}
	\label{fig:Samples}
\end{figure}

The samples from each dimension are subsequently shuffled, to randomize
the order in which each value will be used
(see example on figure \ref{fig:Samples}).
As the samples come from the distributions $D_i$, and are only reordered, their
(marginal) distribution 
will remain that of $D_i$. However, the joint distribution
of the parameters is still not well defined. In particular, this simple shuffling
may result in some of the parameters to be positively or negatively correlated with
each others, which might be undesirable. Some techniques have been developed to
eliminate these correlation terms or to impose different correlations between the
variables, and will be presented on section \ref{ext}.

\begin{figure}[htpb]
	\begin{center}
		\includegraphics[width=200 pt]{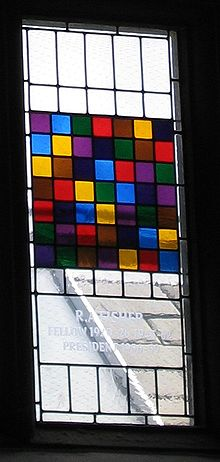}
	\end{center}
	\caption{A stained glass window at the Caius College, Cambridge, 
	showing a full Latin Square. Notice how there is only one occurrence 
	of each color in each row and in each column.}
	\label{fig:glassLS}
\end{figure}

It should be noted that, in the mathematical literature, it is usual to refer 
to a somewhat different object as a Latin Square: this would be a square whose 
sides are divided in $N$ intervals, and is filled with $N$ different symbols, 
such that for each row and column there is exactly one occurrence of each 
symbol, as represented in figure \ref{fig:glassLS}.

\subsection{Algorithms and extensions}\label{ext}
As described above, the LH sampling generates an uniform distribution of samples 
in each parametric dimension. However, there is no guarantee that the correlation 
between two or more parameters will be zero, and the classical algorithm
from McKay, described in the previous section,
usually produces correlations as high as 0.3 between
pairs of factors, which can difficult or even compromise further analyses. In this
section, we will present one algorithm designed to take into
account the correlation between the parameter variables \citep{Huntington98}, using
a single-switch-optimized sample reordering scheme. 
We will present the general case of prescribing a correlation matrix, and will 
also present results for the trivial case of zero correlation terms. Other
methods have been proposed to address this problem \citep{Ye98, Steinberg06, Florian92}, 
including methods that deal with higher-order correlation terms \citep{Tang98}
using orthogonal designs and methods that resort to stochastic optimization based on 
simulated annealing \citep{Vovrechovsky09}. These methods, however, either impose severe 
restrictions on the number of samples that must be chosen or are too
computationally intensive.

In order to obtain the samples with prescribed correlation terms, we define 
as $C_{i,j}$ the 
desired $m \times m$ correlation matrix between the variables $x_i$ and $x_j$,
and denote by $C_{i,j}^*$ the current correlation between $x_i$ and $x_j$.

The next step is done iteratively for each parameter dimension, starting with 
the second one. Suppose that the method has already been applied to $i=1,2,....,l-1$, 
and we will apply it to $i=l$. The square sum of the errors in the correlations 
between $x_l$ and the anterior parameters is given by

\begin{equation}
	E = \sum_{k=1}^{l-1} \left( C_{l, k} - C_{l,k}^* \right) ^2
	\label{Error}
\end{equation}

Afterwards, we calculate, for each pair of values sampled from the parameter 
dimension $l$, what would be the error in the correlation if they were switched.
The pair that corresponds to the greater error reduction is then switched, and
the procedure is repeated iteratively until the error is acceptably small.

\subsubsection{Note on the existence of solutions}
The problem of generating a sample with specified marginal distributions and correlation
terms is tightly connected to the problem of generating samples from a multivariate
distribution. This class of problems is very complex, and has received limited
attention from the probabilist community, with the recent paper describing the
exact construction of all feasible bivariate exponential distributions being considered
a significant theoretical advance \citep{Bladt10}.

One surprising result by \cite{Hoeffding1940} ({\em apud} \citep{Dukic13}) is that
the specified distribution function need not exist. For any pair of marginal distributions
$D_1$ and $D_2$, there exist a maximum and a minimum correlation coefficients,
$\rho^{+}$ and $\rho^{-}$ such that there exists a joint probability distribution $\bu{D}$
with specified marginal distributions and correlation. For some marginal distributions,
such as the Gaussian, these values are $\rho^+=1$ and $\rho^-=-1$, so that any correlation
term results in a valid distribution. On the other hand, if both $D_1$ and $D_2$ are
exponential distributions with parameter $\lambda=1$, the values change to $\rho^+=1$ and
$\rho^-=1-\pi^2/6 \approx -0.65$. 

In short, finding a Latin Hypercube with exponential marginal distributions and
at least one correlation term of, for example, $-0.9$ is impossible, no matter which 
algorithm is employed. 
However important this result is for the theory of probability,
it may not have strong consequences in practical applications. If the marginal distributions
and correlation terms are chosen after theoretical considerations or real world data, the
impossible probability distributions will not be attempted.

For recent developments in this area, see the work of \cite{Dukic13} and \cite{Huber14}.

\subsection{Stochastic models} %TODO menos handwaving
When dealing with stochastic models, like several relevant individual based 
models (IBM), the questions presented become complicated by the fact that 
running the same model with exactly the same parameters might wield largely 
different results, both quantitative and qualitatively. In this scenario,
we must be able to differentiate the variation in responses due to the variation
of the parameters with the variation in response due to stochastic effects.

We will refer to the variation due to the input parameters as
the epistemic uncertainty. This uncertainty arises from the fact that we do
now know what are the correct values for a given parameter in a given natural
system, and is related to the probability distributions $D_i$, presented 
in section \ref{PSE}. The variation in the behavior of the model which is
caused by stochastic effects, for a fixed set of parameters,
is called stochastic uncertainty, and is inherent
to the model. 

It is important to note that the two uncertainty components are impossible
to disentangle in general stochastic models. This has prevented the general
analysis of such models until recently. In recent years, studies have shown
that the important parameters and their effects can be correctly identified
by running such models repeatedly for the same input variables and then
averaging the output \citep{Segovia04}, given that
the following conditions are respected:

\begin{itemize}
	\item Sample sizes should be large, relative to the stochastic
		uncertainty.
	\item The output values should be unimodal, that is, the output values
		for a given parameter choice should be clustered around a 
		central value.
	\item The correct analysis tools should be used (as will be
		discussed on session \ref{QuantAnal}).
\end{itemize}

\subsection{Measuring the concordance with increasing sample size}
\label{SBMA}
We will now turn our attention to the problem of determining the optimal
number of model runs we should apply in order to provide a good estimate
of which are the relevant parameters for a given model.
One way of proceeding is by systematically increasing the number $N$ of
model runs and applying any of the sensitivity analysis techniques, which
will be discussed on the following section. If the analyses indicate 
similar results for consecutive runs, we can presume that 
increasing the sample size will not yield major changes to the results.

All of the sensitivity analyses present us with a list of the parameters 
that have most influence in the model output. By comparing the resulting 
lists from two experiments, we can decide to stop increasing $N$ when
the lists are sufficiently similar. Our problem then is to determine
how similar are two vectors of ranks. In principle, we could
apply any distance function to those vectors. However, consider that
3 analyses indicated that the order of the most influential parameters is:
\begin{center}
\begin{tabular}[h!]{c c c c c c c c}
		\hline
		H1 &=& 1& 2& 3& 4& 5& 6\\
		H2 &=& 1& 2& 3& 6& 4& 5\\
		H3 &=& 2& 3& 1& 4& 5& 6\\
		\hline
\end{tabular}
\end{center}
By using standard distances (like Spearman's rho or Kendall's tau),
we will see the same difference between H1 and H2 and between H1 and H3.
On the other hand, in the context of determining the most influential
parameters, we would be inclined to see H1 and H2 as more similar than
any of them to H3, as the first two preserve the ordering of the three
first parameters.

Iman and Conover proposed a correlation coefficient for this problem
called Top-Down Correlation Coefficient \citep{ImanConover87}, which is
based on Savage Scores. This coefficient, know as TDCC, 
was extensively used for sensitivity analyses \citep{Marino08}. 
Another measure of concordance proposed more recently is the Symmetrized
Blest Measure of Association (SBMA) \citep{Genest03}.
Recent research suggests that estimates for SBMA 
produce a smaller standard error than TDCC without the assumption that 
there is no correlation between the variables \citep{Maturi10}.
Thus, we propose using SBMA as a measure of concordance between analyses
from different sample sizes. Defining the ranks from the first sample as
$R_i$ and the ranks for the second sample as $S_i$, the estimator for the
SBMA is:
\begin{equation}
		\xi_n = - \frac{4n+5}{n-1}+\frac{6}{n^3-n} %
		\sum _{i=1}^n R_iS_i \left( 4- \frac{R_i+S_i}{n+1} \right)
		\label{eqSBMA}
\end{equation}

For the techniques that may output negative values, for instance negative
correlations, the SBMA must be applied on the absolute values. Otherwise,
the parameters which present strong negative effects will be ranked very low,
and will not be taken into account by the SBMA.

We will apply SBMA for the PRCC technique (discussed on section \ref{QuantAnal})
on the example on section \ref{Leslie}.

\section{Quantitative output analysis}\label{QuantAnal}
\subsection{Uncertainty analysis}\label{UA}
The first question we would like to answer, in the context of quantitative 
analysis, is what is the probability distribution of the response variable $y$
given that we know the join probabilities of the input parameters $\bu{x}$ 
(see definitions in section \ref{PSE}), which is the subject of uncertainty
analysis \citep{Helton03}. 

This can be done by fitting a density curve to the output $y$ or an empiric
cumulative distribution function (ecdf). If there is any theoretical reason
to believe that the distribution of $y$ should follow one given distribution,
it is possible to fit this function to the actual output data and estimate
the distribution parameters. If the joint distribution of the input parameters
correspond to the actual probability of some natural system to exhibit some
given set of parameter values (as opposed to the case where we have no 
biologically 
relevant estimates for some parameters), the estimate represented by the
density and ecdf functions approaches the actual distribution that the
variable $y$ should present in nature. This functions may be used, for example,
to provide confidence intervals on the model responses.

However, this is only the case when the input variables are uncorrelated or
when enough correlation terms have been taken into account. \cite{Smith02}
provides an example where ignoring the correlation terms 
leads to inaccuracies on the estimation of confidence intervals.

The next reasonable step is to construct and interpret scatterplots relating
the result to each input parameter. These scatterplots may aid in the visual
identification of patterns, and although they cannot be used to prove 
any relationship between the model response and input, they may direct the
research effort to the correct analyses. There are extensive reviews of the
use of scatterplots to identify the important factors and emerging patterns
in sensitivity analyses \citep{Kleijnen99}.

We will present here some quantitative analyses tools, aimed at identifying
increasingly complex patterns in the model responses. It should be stressed
that no single tool will capture all the relations between the input and
output. Instead, several tools should be applied to any particular model.

\subsection{Sensitivity analysis}
The question of ``what is the effect of some combination of parameters to
the model output'' may be answered by testing the relation between
the parameters and outputs. 
There are extensive reviews about detecting these relations after
generating samples with Latin Hypercubes \citep{Marino08, Kleijnen99}, 
so we will give just a brief overview. We will first note that
the methods used must take into account the variation of all the 
parameters. For example, instead of calculating the correlation between
the result and some parameter, partial correlation coefficients should
be used, which
discount the effect of all other parameters. 

The classical approach to the sensitivity analysis, based on the frequentist
school of hypothesis testing, consists in classifying the relations
between the results and the input parameters, in order of increasing complexity, as:

\begin{itemize}
	\item Linear relation, which can be tested with the Pearson partial correlation coefficient.
		Is is usual to test the significance of this linear relation by a t-test 
		\citep{Freedman07}.
	\item Monotonic relation, which can be tested with the Spearman partial correlation coefficient,
		also referred to as Partial Rank Correlation Coefficient, or PRCC.
		This measure is a robust
		indicator of monotonic interactions between $y$ and $x_i$, and is subject
		to significance testing \citep{Marino08}.
	\item Trends in central location, for which the Kruskal-Wallis test may be applied \citep{Kleijnen99}.
	\item Trends in variability, for which the FAST method and Sobol' indexes may be used in order to partition
		the model variability \citep{Archer97, Saltelli99, Saltelli04}.
\end{itemize}

Subsections \ref{linear} to \ref{FAST} will provide some
mathematical background for each method, and section \ref{Leslie} 
will present examples of use of those tests. We should stress here that
the application of one method is not enough to draw conclusions about 
the relations between the input and output variables, as these techniques
test different hypotheses, and have different statistical powers. Instead,
every model should be analyzed by a combination of techniques, preferably
one for each category outlined here.

\subsubsection{Linear relation}\label{linear}
Under the hypothesis of independence between the central location and
dispersion of the model responses, 
the most straightforward relationship between $y$ and $x_i$ is the linear, 
represented by $y \sim x_i$. This is case if, every time $x_i$ is increased, 
$y$ increases by approximately the same amount.
The Pearson correlation coefficient is the commonly used measure to test for
a linear correlation:

\begin{equation}
	\rho_{yx_i} = \frac{\sigma_{yx_i}}{\sigma_y\sigma_{x_i}}
	\label{PearsonRho}
\end{equation}

Where $\sigma_a$ is the variance of $a$ and $\sigma_{ab}$ is the covariance 
between $a$ and $b$. The correlation coefficient is a measure of the predicted
change in $y$ when $x_i$ is changed one unit, relative to its standard
deviations, and, as such, approaches $\pm 1$ when there is a strong
linear relation between the variables. The square of $\rho$, usually written as
$R^2$, measures the fraction of the variance in the output that can be accounted
for by a linear effect of $x_i$.
Is is usual to test the significance of this linear relation by a t-test 
\citep{Freedman07}.

Other than examining the individual relationships between the parameters and 
the output, we can investigate the joint effect of several $x_i$, as
$y \sim x_1 + x_2 + \cdots + x_m$. In this case, the multiple $R^2$ represent
the fraction of the variance on the output due to linear effects of all 
the $x_i$ considered.

However, a measure of $\rho$ close to zero does not mean that
no relationship exists between $y$ and $x_i$ - for instance, $x^2 + y = 1$, 
$x \in [-1,1]$ presents $\rho = 0$, so clearly other methods might be needed.

The Partial Correlation Coefficient (PCC) between $x_i$ and $y$ is the measure
of the linear effect of $x_i$ on $y$ after the linear effects of the remaining
parameters have been discounted. In order to calculate the PCC, first we fit
a linear model of $x_i$ as a function of the remaining parameters:

\begin{equation}
	\hat{x}_i \sim x_1 + x_2 + \cdots + x_{i-1} + x_{i+1} + \cdots + x_m
	\label{PCChatx}
\end{equation}

A corresponding model is done with $y$:

\begin{equation}
	\hat{y} \sim x_1 + x_2 + \cdots + x_{i-1} + x_{i+1} + \cdots + x_m
	\label{PCChaty}
\end{equation}

The PCC is calculated as the correlation between the residuals of these two
models:

\begin{equation}
	PCC(y, x_i) = \rho \left( (y - \hat y), (x_i - \hat{x}_i) \right)
	\label{PCC}
\end{equation}

\subsubsection{Monotonic relation}\label{PRCC}

Let us refer to each value of $y$ as $y_k$ and each value of $x_i$ as 
$x_{ik}$. The rank transformation of $y$, 
represented by $r(y_k)$ can
be found by sorting the values $y_k$, and assigning rank 1 to the smallest, 2 
to the second smallest, etc, and $N$ to the largest. The rank of $x_{ik}$,
$r(x_{ik})$, can be found in a similar way.

If there exists a strictly monotonic relation between $y$ and $x_i$, that is,
if every time $x_i$ increases, $y$ either always increase or always decreases
by any positive amount, it should
be clear that the ranks of $y$ and $x_i$ present a linear relationship: 
$r(y) \sim r(x_i)$.

The correlation between $r(y)$ and $r(x_i)$ is called the Spearman 
correlation coefficient
$\eta_{yx_i}$. The same analyses presented on section \ref{linear} can also be
applied for the rank transformed data, including significance testing and 
multiple regression.

If the procedure described to calculate the PCC is followed on rank transformed
data, that is, if $y$ and $x_i$ are rank transformed and fitted as linear models
of the remaining parameters, the correlation between the residuals is called
PRCC, or Partial Rank Correlation Coefficient. This measure is a robust
indicator of monotonic interactions between $y$ and $x_i$, and is subject
to significance testing \citep{Marino08}. This measure will 
perform better with increasing $N$.

\subsubsection{Trends in central location}\label{Kruskal}
Even if the relation between $y$ and $x_i$ is non monotonic, it may be 
important and well-defined. The case in which $y \sim x_i^2$, $x_i \in (-1,1)$
is a common example. This relation may be difficult to visualize, and sometimes
may not be expressed analytically. In these cases, the Kruskal-Wallis rank
sum test may be used to indicate the presence of such relations 
\citep{Kleijnen99}.

In order to perform the test, the distribution of $x_i$ must be divided 
into a number $N_{test}$ of disjoint intervals. The model response $y$ is then
grouped with respect to these intervals, and the Kruskal-Wallis test is used
to investigate if the $y$ values have approximately the same distribution
in each of those intervals. A low p-value for this test indicates that the
mean and median of $y$ is likely to be different for each interval considered,
and thus that the parameter $x_i$ have a (possibly non monotonic) relationship
with $y$.

\begin{figure}[htbp]
	\begin{center}
\includegraphics{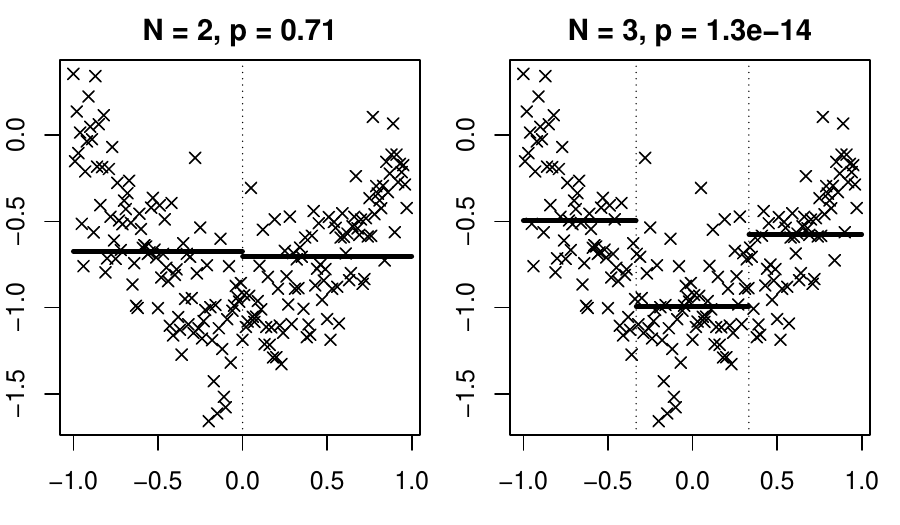}
	\end{center}
	\caption{Example of application of the Kruskal-Wallis test on the
	same data set, which present a strong quadratic component, by
	dividing the range in 2 intervals (right) or 3 intervals (left).
	The dashed lines are the divisions between the intervals, and
	the strong horizontal lines are the sample means for each 
	interval.}
	\label{fig:Kruskal}
\end{figure}

The number of intervals $N_{test}$ is not fixed as any ``magical number'', and
may have a large impact on the test results. It is then recommended that 
this test should be repeated with different values to obtain a more 
comprehensive picture of the interactions between $x_i$ and $y$ (fig. \ref{fig:Kruskal}).

\subsubsection{Trends in variability}\label{FAST}
Other than the central tendency of the results, their dispersal may be 
dependent on the input parameters. A classical approach which may be used
to test whether the dispersal of the output is related to any input
parameter is to divide the distribution of $x_i$ 
into a number $N_{test}$ of disjoint intervals and group the model response
$y$ with respect to these intervals, as done on the Kruskal-Wallis test.
In this case, the ANOVA F statistic can be used to test for equality 
between the $y$ conditional to each class \citep{Kleijnen99}.

A similar approach, which we will use, is to employ the eFAST indexes
\citep{Saltelli99, Saltelli04},
which is a variance decomposition method based on the FAST and
Sobol' indexes. While the Sobol' indexes were described in 1969, in Russian,
FAST was developed by Cukier et al. in 1973, and both are identical in all
but one computation \citep{Archer97}. 

These methods estimate what fraction of the output variance can be explained
by variation in each parameter $x_i$, which is called the {\em first-order 
sensitivity of $x_i$} or {\em main effect of $x_i$}. The method estimates
as well the fraction which is explained by 
the higher-order interactions between $x_i$ and all other parameters.
The sum of all terms related to $x_i$ is called the {\em total-order
sensitivity of $x_i$}. 

The eFAST method estimates the main effect of each parameter by choosing 
a periodic function $f_i(x_i)$ for each parameter, where the 
frequency $\phi_i$ of
each function is distinct, and should, in theory, be incommensurable.
Each of this functions is sampled $N_s$ times, and a Fourier analysis
is applied to the model output. The Fourier coefficients at each frequency
$\phi_i$ is related to the main effect of the variable $x_i$. The total
order sensitivity of $x_i$ is then calculated as the fraction of the
variation which is not explained by the complimentary of $x_i$ (that is,
all parameters but this one).

There are two things that should be noted here about eFAST indexes. The
first one is that the eFAST calculation does not involve the LHS sampling
scheme, and may require more model evaluations. Also, this method produces
small positive total-order sensitivity estimates even for parameters 
which do
not play any role on the model output, as many numeric approximations are
involved. 

\subsection{Bayesian alternatives}
The Bayesian view of statistics present some alternatives to the 
techniques outlined in the previous sections. Beven and Binley described
a procedure for the Bayesian updating of probabilities called GLUE,
for Generalized Likelihood Uncertainty Estimation\citep{Beven92}. While
the previously discussed methodologies are appropriate for purely exploratory
analyses, the GLUE method is suited for problems in which one or more of 
the parameters of the model require calibration using the object of prediction.
It is based on the notion that, for a given model response, there is always 
a set of models that will recreate it. This set is called {\em equifinal}.

Despite the widespread use and recognition that the GLUE method has received,
it is subject to criticisms by not being formally Bayesian, and
formal Bayesian approaches have been developed\citep{Vrugt09}.
Another approach, based on the Metropolis algorithm, is provided by 
\cite{Kuczera98}.

\section{Case study 1: a structured model of {\em Euterpe edulis} populations}\label{Leslie}
\subsection{Model description}
In this section, we demonstrate the uses and advantages of the methods
outlined in the previous sections by performing sensitivity analyses
on a density-dependent model of the tropical palm {\em Euterpe edulis}
(commonly known as palmito ju\c cara). All the data used here
was extracted from Silva Matos {\em et al.} paper \citep{SilvaMatos99},
which compared a density independent matrix model of population growth
with a density dependent model in which the recruitment of seedlings was
affected by the number of seedings and adult trees. Silva Matos
provided results, sensitivities and elasticities for the density independent 
model that can be compared to our findings, and results for mean and maximum
values of the density dependent model - but unfortunately, their methods
did not allow for a full sensitivity analysis of the density dependent
model.

We have used the \R language to perform the sampling and analysis, 
with the ``pse'' package, which implements the tools described
in the previous sections. We have also used code from
the ``sensitivity'' package, which implements 
PRCC analysis (see section \ref{PRCC}) and eFAST analysis (see 
section \ref{FAST}), among
others. The ``pse'' package also implements the Huntington \& Lyrintzis' 
algorithm to generate zero correlation between LHS samples (see
section \ref{LHS}). All code used is freely available on the web.

The models analyzed are based on a Lefkovitch matrix with seven size
classes. The matrix used on the density-independent model is

\begin{equation}
		A = \left[
		\begin{array} {ccccccc}
				P_1 &   0 &   0 &   0 &   0 &   0 & F_7 \\
				G_1 & P_2 &   0 &   0 &   0 &   0 &   0 \\
				  0 & G_2 & P_3 &   0 &   0 &   0 &   0 \\
				  0 &   0 & G_3 & P_4 &   0 &   0 &   0 \\
				  0 &   0 &   0 & G_4 & P_5 &   0 &   0 \\
				  0 &   0 &   0 &   0 & G_5 & P_6 &   0 \\
				  0 &   0 &   0 &   0 &   0 & G_6 & P_7 
		\end{array}
		\right]
		\label{LefMatrix}
\end{equation}

Here, $P_i$ is the 
probability of a tree surviving and remaining in the same class
(stasis), $G_i$
is the probability of a tree surviving and growing to the next class,
and $F_i$ is the number of offspring produced per reproductive palm.

The dominant eigenvalue of this matrix is related to the predicted population
growth rate. We considered the dominant eigenvalue for this matrix as
the model output, as usually done on this modeling approach.

The density dependent model used the same matrix, but now the growth
term of the first size class represented a decreasing function of the population
density:

\begin{equation}
		G_1 = \frac{ G_m }{1+ a N_1} \exp \left(- \frac{z}{\rho} N_7 \right)
		\label{G_1}
\end{equation}

Here, $N_1$ and $N_7$ represent the number of seedlings and adults per patch.
The parameters $G_m$ and $a$ represent the maximum transition 
rate at low densities and the strength of reduction in $G_1$ with 
increasing seedling densities. The remaining parameters $z$ and $\rho$
represent the crown area of an adult tree and the plot size (which is 
fixed as 25$m^2$), and their ratio is related to the reduction of
recruitment due to the presence of adults, due to the fact that few
seedlings are able to grow underneath the canopy of an adult. 

As this model does not produce any static matrix, it is not meaningful to
calculate any eigenvalue. Instead, the total population corresponding to
the stable population distribution was used as model output.

A na\"ive approach to estimating the parameter sensitivities of this model would 
use the stasis, growth and fecundities given. However, this would yield
erroneous results,
as the probabilities of stasis and growth for a given class are not independent, as
$P_i+G_i \leq 1$ for all classes. As discussed on section \ref{ps}, we need to
use an alternative parametrization for this model.

We will represent by $s_i$ the probability of survival for each class, calculated
as $s_i = P_i+G_i$, and by lowercase $g_i$ the probability of growth, calculated
as $g_i = (s_i - P_i) / s_i$. Using the notation for complementary probabilities
$\overline{g_i} = 1-g_i$, we can now write the Lefkovitch matrix as:

\begin{equation}
		A = \left[
		\begin{array} {ccccccc}
				s_1 \cdot \overline{g_1} &   0 &   0 &   0 &   0 &   0 & F_7 \\
				s_1 \cdot g_1 & s_2 \cdot \overline{g_2} &   0 &   0 &   0 &   0 &   0 \\
				  0 & s_2 \cdot g_2 & s_3 \cdot \overline{g_3} &   0 &   0 &   0 &   0 \\
				  0 &   0 & s_3 \cdot g_3 & s_4 \cdot \overline{g_4} &   0 &   0 &   0 \\
				  0 &   0 &   0 & s_4 \cdot g_4 & s_5 \cdot \overline{g_5} &   0 &   0 \\
				  0 &   0 &   0 &   0 & s_5 \cdot g_5 & s_6 \cdot \overline{g_6} &   0 \\
				  0 &   0 &   0 &   0 &   0 & s_6 \cdot g_6 & s_7 
		\end{array}
		\right]
		\label{LefMatrixReparam}
\end{equation}

The models have, respectively, 14 and 16 parameters. All analyses have 
been done with mean and standard deviation calculated from Silva Matos paper,
assuming a normal distribution of parameters truncated at the $[0,1]$ interval
for probabilities, and on $[0,+\infty)$ for the other parameters.
In the case of the density dependence
parameters $Gm$, $z$ and $a$, only the mean estimate was given on the paper,
so conservative values were used for the standard deviations.

\begin{figure}
\includegraphics{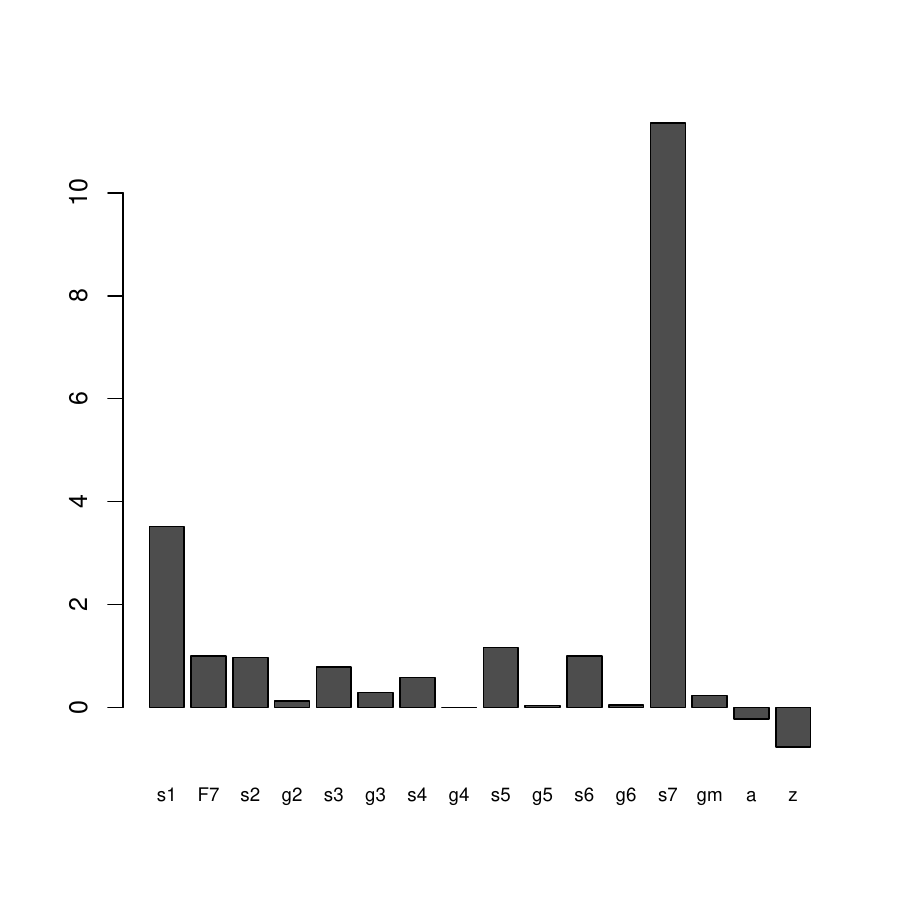}
	\label{fig:analytical}
	\caption{Analytical elasticities for the density-dependent model}
\end{figure}

We have used the methods described in \cite{Caswell08, Caswell09, Caswell10}
to estimate the analytical elasticities of the model in fig. \ref{fig:analytical}.

\subsection{Results}
First, we have generated Latin Hypercubes consisting of all relevant variables
for each model. Then, the models were run for each combination of parameters. 
By using the SBMA measure of concordance (section \ref{SBMA}), we have determined that the sample size
required for the density independent model is approximately 100, 
and between 300 and 500 for the density dependent (Table \ref{tabprcc}).

\begin{figure}[htpb]
		\begin{center}
\includegraphics{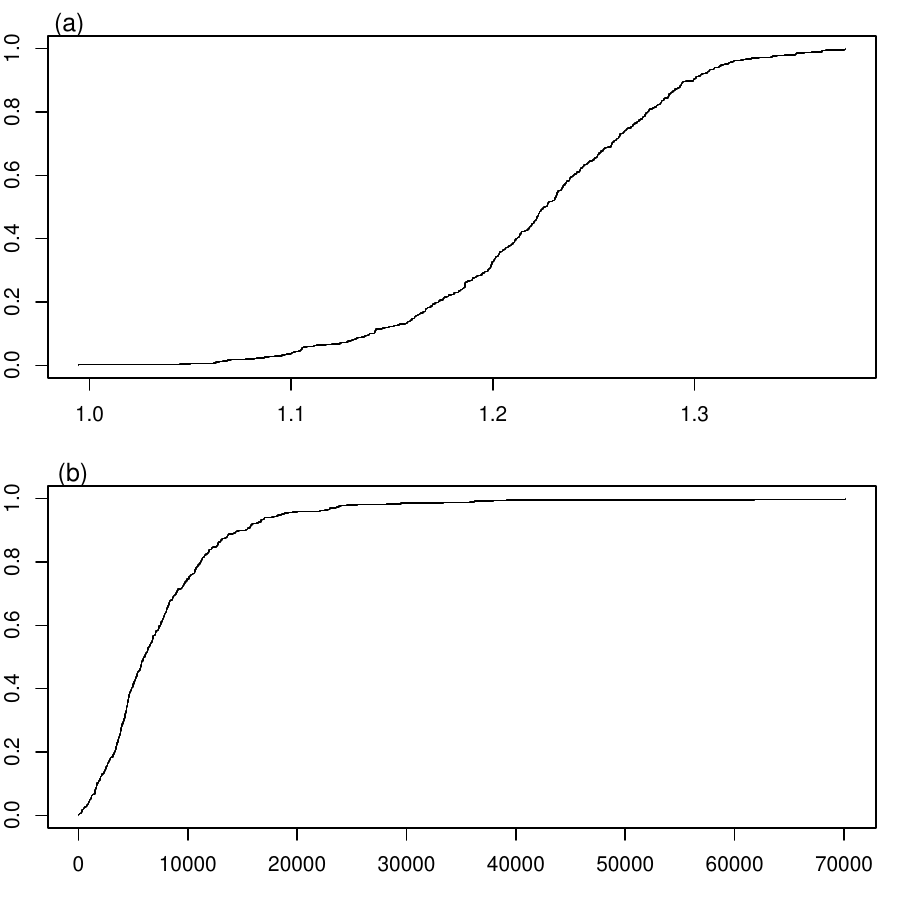}
		\end{center}
		\caption{Empirical cumulative distribution functions (ecdf) for the
		density independent (a) and density dependent (b) models of population 
		growth. In (a), the x axis represents the dominant eigenvalue, and the
		population is viable if x > 1. In (b), the x axis represents the total
		equilibrium population, and the population is viable if x > 0.}
		\label{fig:ecdf}
\end{figure}

% latex table generated in R 3.2.1 by xtable 1.7-3 package
% Tue Jul  7 00:13:40 2015
\begin{table}[ht]
\centering
\begin{tabular}{rlrr}
  \hline
 & Size & Independent & Dependent \\ 
  \hline
1 & 50-100 & 0.87 & 0.51 \\ 
  2 & 100-200 & 0.86 & 0.49 \\ 
  3 & 200-300 & 0.92 & 0.89 \\ 
  4 & 300-400 & 0.93 & 0.82 \\ 
  5 & 400-500 & 0.91 & 0.87 \\ 
   \hline
\end{tabular}
\caption{Comparison of PRCC analyses by sample size for both models} 
\label{tabprcc}
\end{table}
If we presume that the data collected is representative of our knowledge about each
of these parameters, the probability distribution of the model responses can 
be seen as the probability that the real population of palms exhibit each
value of the model output.
Figure \ref{fig:ecdf} shows these distributions, which suggest that
the population is viable for the vast majority of parameters values in the 
parameter space considered. Also, the  
$\lambda$ calculated from the density-independent model 
(mean 1.22, standard deviation 0.06), 
is very close to the value found by Silva Matos (mean $1.24 \pm 0.06$ se). 
Considering the density-dependent
model, the median stable population predicted 
(6028
trees in each $25m^2$ plot), 
is comparable to, although higher than, the population  
actually measured by the study ($1960 \pm 560$ trees per plot, mean and 
sd calculated over three years).

\begin{figure}[htpb]
		\begin{center}
\includegraphics{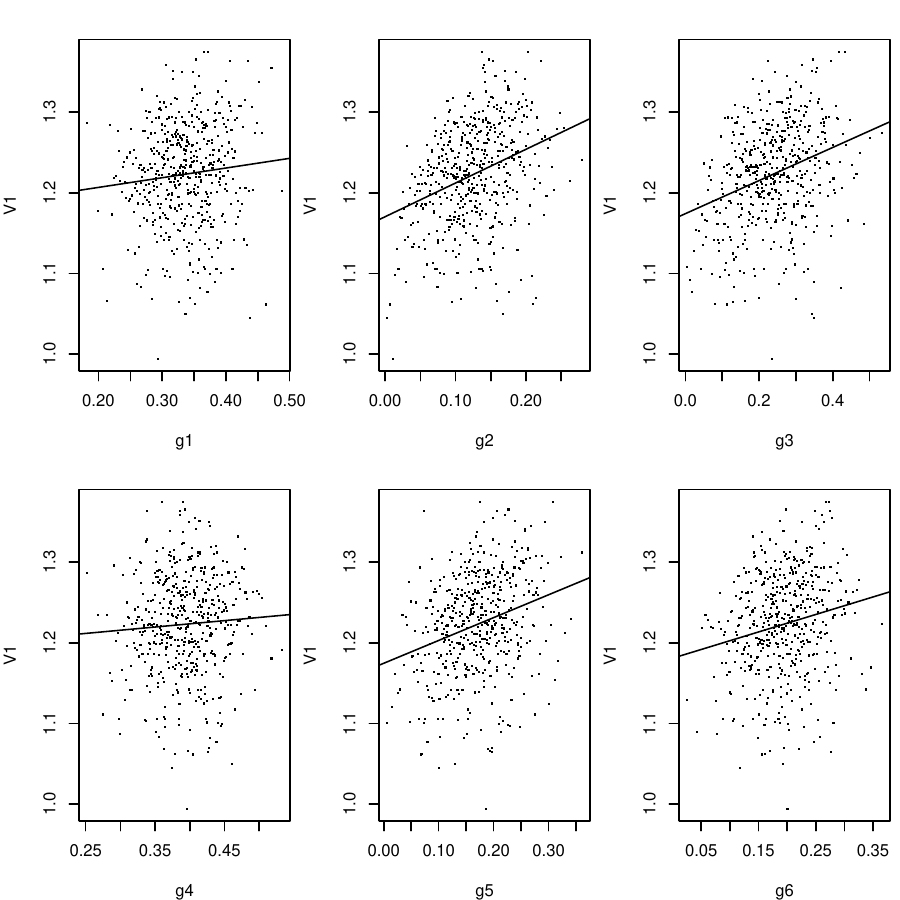}
		\end{center}
		\caption{Scatterplots relating the value of the input parameters
		of growth to the $\lambda$ calculated
		the output for the density independent model.}
		\label{fig:iCorr1}
\end{figure}

\begin{figure}[htpb]
		\begin{center}
\includegraphics{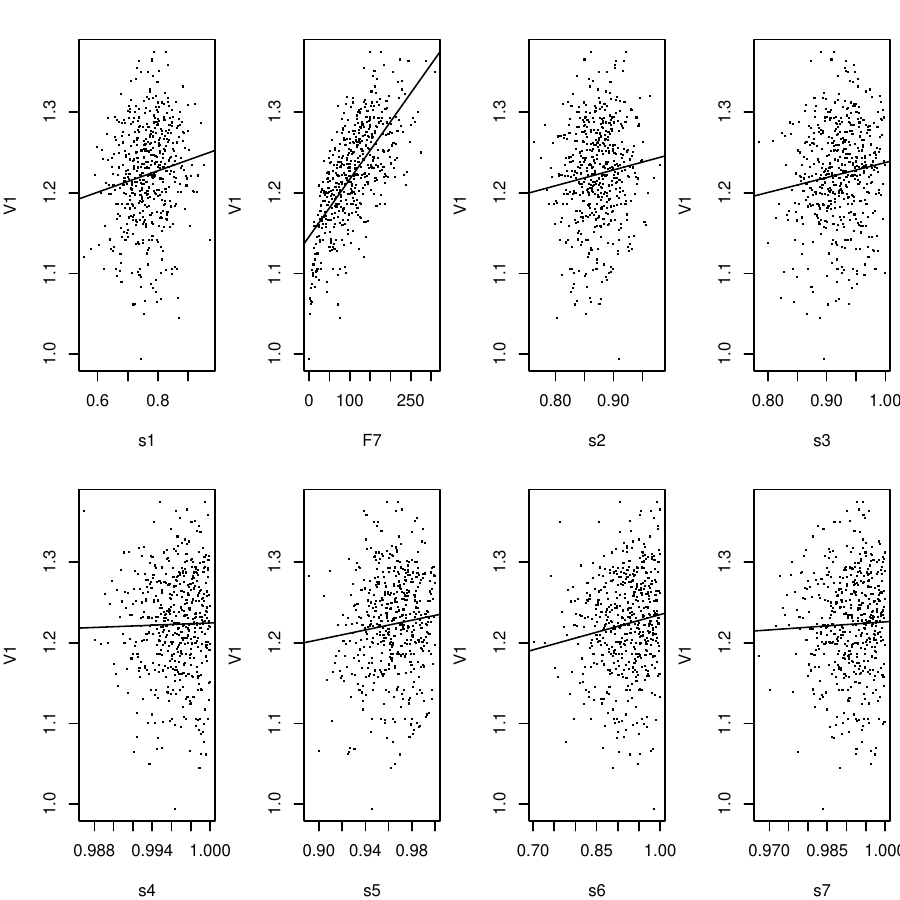}
		\end{center}
		\caption{Scatterplots relating the value of the input parameters
		of survival and fecundity to the $\lambda$ calculated
		the output for the density independent model.}
		\label{fig:iCorr2}
\end{figure}

\begin{figure}[htpb]
		\begin{center}
\includegraphics{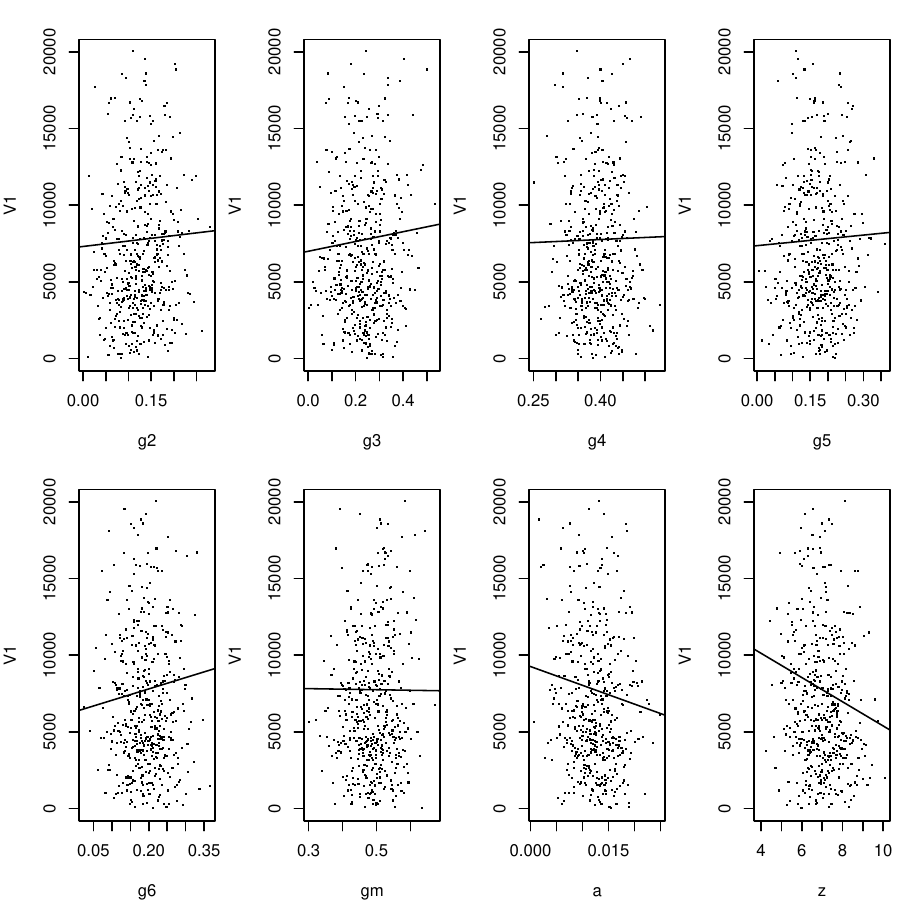}
		\end{center}
		\caption{Scatterplots relating the value of the input parameter 
		of growth and density-dependence to
		the output for the density dependent model.}
		\label{fig:dCorr1}
\end{figure}

\begin{figure}[htpb]
		\begin{center}
\includegraphics{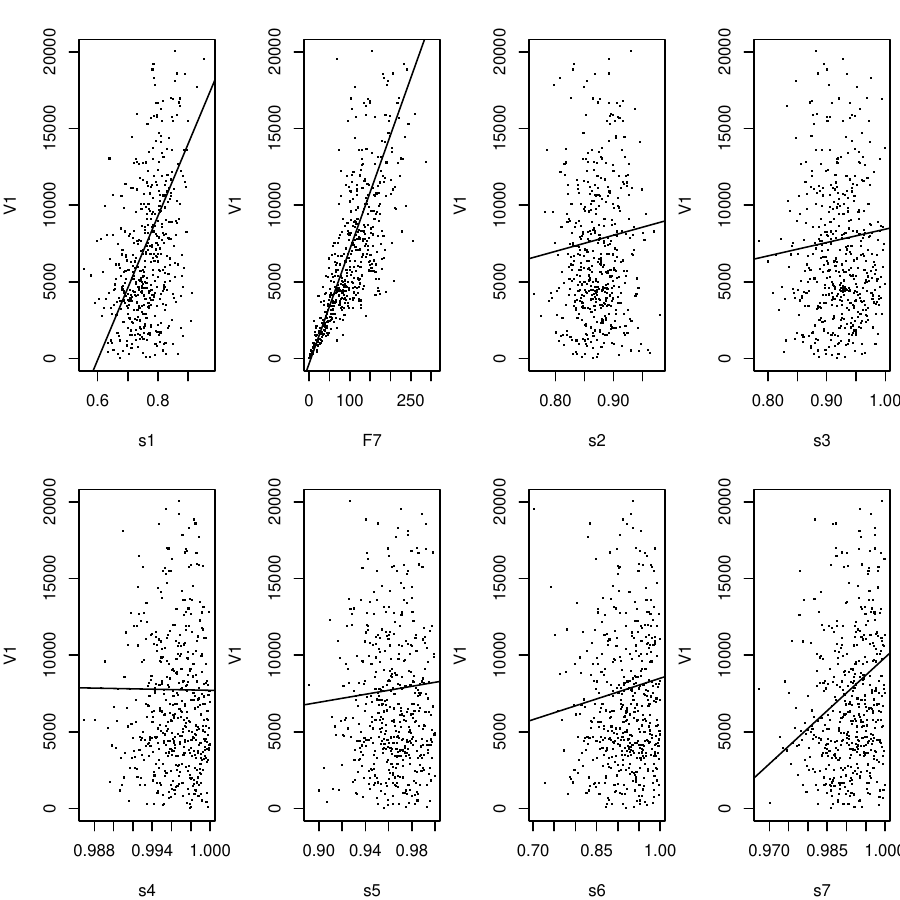}
		\end{center}
		\caption{Scatterplots relating the value of each input parameter of
		survival and fecundity to
		the output for the density dependent model.}
		\label{fig:dCorr2}
\end{figure}

We have generated scatterplots between the result from the models and each 
independent parameter, in order to visually identify the relations
between the inputs and outputs (figs. \ref{fig:iCorr1} to \ref{fig:dCorr2}). 
It is clear from these scatterplots that the fecundity plays a major role
on the population dynamics, and may be involved in non-linear interactions.
Also, growth probabilities ($g_i$) have a 
greater impact on the model output than survival ($s_i$) on the 
density-independent model. This is to be
contrasted with Silva Matos results, which show all of the elasticities
to be approximately equal for all parameters.
In the density-dependent model, the patterns are much more complex. 
Survival parameters seem to be more influent than growth, and
the parameters reducing the recruitment ($a$ and $z$) show a clear
negative effect on the population size. However, there is evidence now
for non-linear effects of the parameters, in particular $s_1$ and $F_7$.


\begin{figure}[htpb]
		\begin{center}
\includegraphics{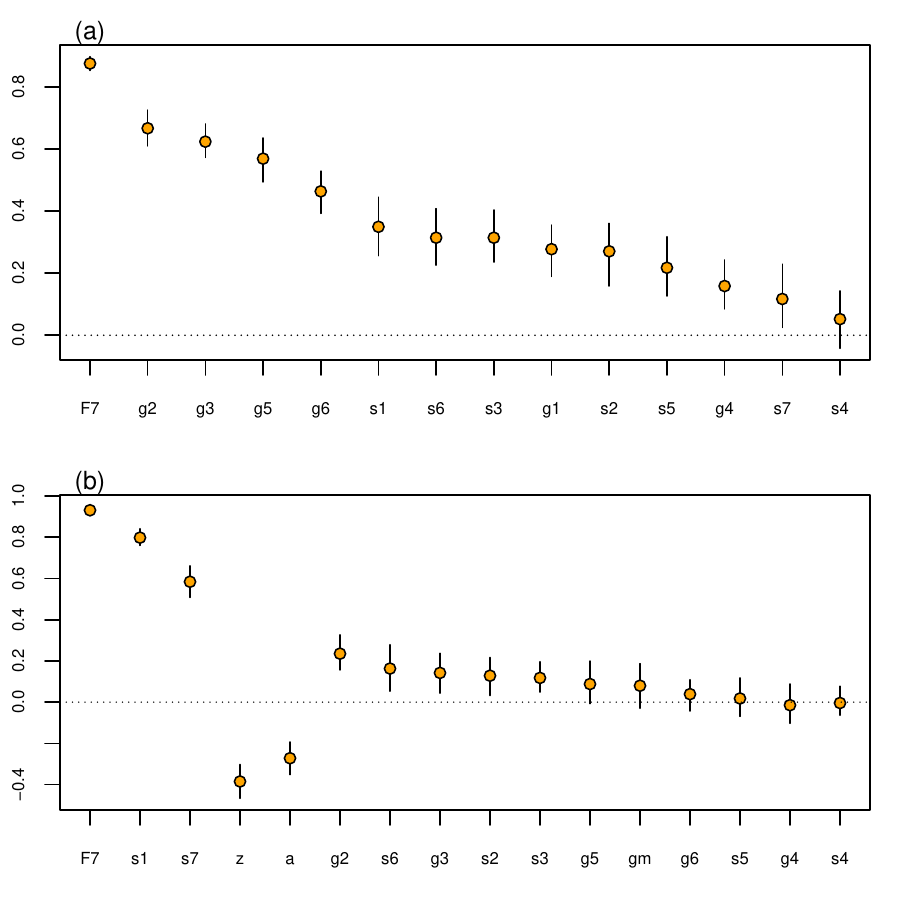}
		\end{center}
		\caption{Partial Rank Correlation Coefficients for the density independent (a)
		and density dependent (b) models. The bars are confidence intervals, generated
		by bootstrapping 1000 times}
		\label{fig:PRCC}
\end{figure}

These scatterplots show very high dispersion of values, mostly due to the fact that
all parameters are being varied between runs. In order to investigate the effect of 
each parameter on the outputs discounting the effects of the others, we analyse the
Partial Rank Correlation Coefficient (PRCC, fig. \ref{fig:PRCC}). The PRCC analysis for the density
independent model is in agreement with our previous expectations, with $F_7$ being
the most influential parameter, followed by growth probabilities. Survival probabilities
follow with low correlations. The density dependent model presents us with some
surprises, as now the survival parameter for the smallest and largest size classes
jumped to occupy the second and third largest positive PRCCs. The remaining parameters
follow the new parameters $a$ and $z$, which are strongly negatively correlated with the output.

It is interesting to contrast these results with the analytical analyses, presented on fig. 
\ref{fig:analytical}. While all the elasticities have the same sign, and a comparable order, the most
conspicuous difference between the two is the high importance given to $s_7$ (the survival of adults)
given by the analytical analysis. This might be interpreted by noticing that this parameter is the one 
for which the collected data leaves the smallest margin of uncertainty.

\begin{figure}[htpb]
		\begin{center}
\includegraphics{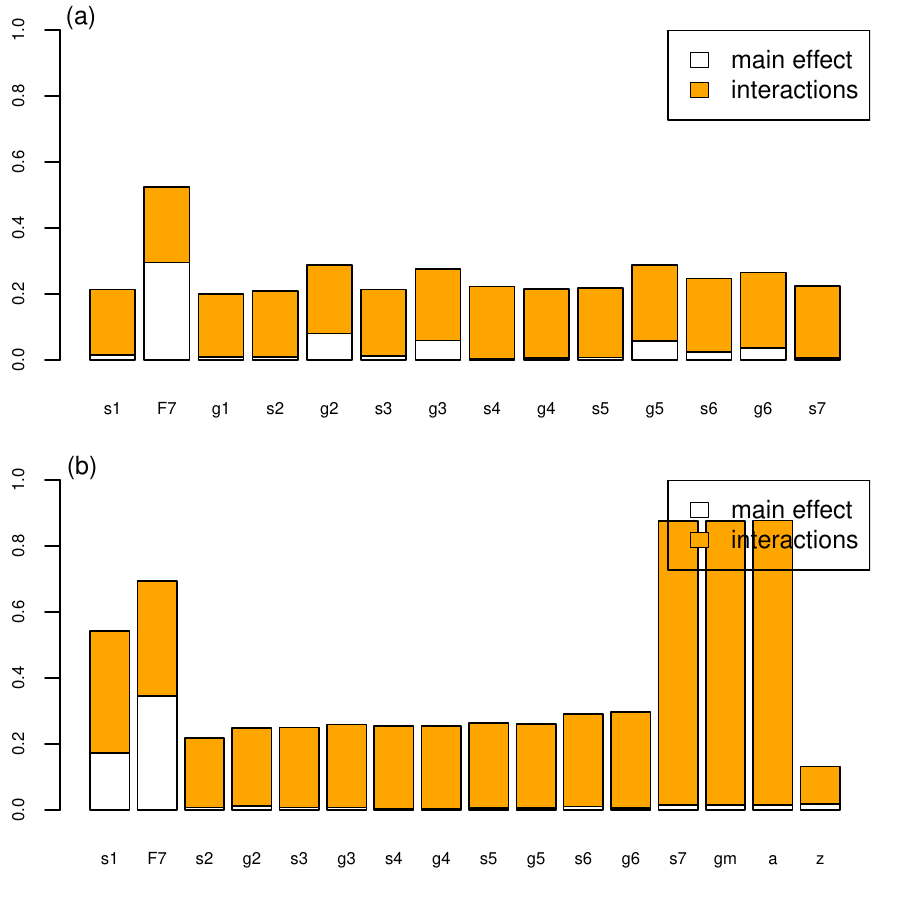}
		\end{center}
		\caption{eFAST analysis for the density independent (a)
		and density dependent (b) models. The bars represent the first and total order
		estimates for the sensitivity of each parameter in the model output.}
		\label{fig:fast}
\end{figure}

\begin{figure}[htpb]
		\begin{center}
\includegraphics{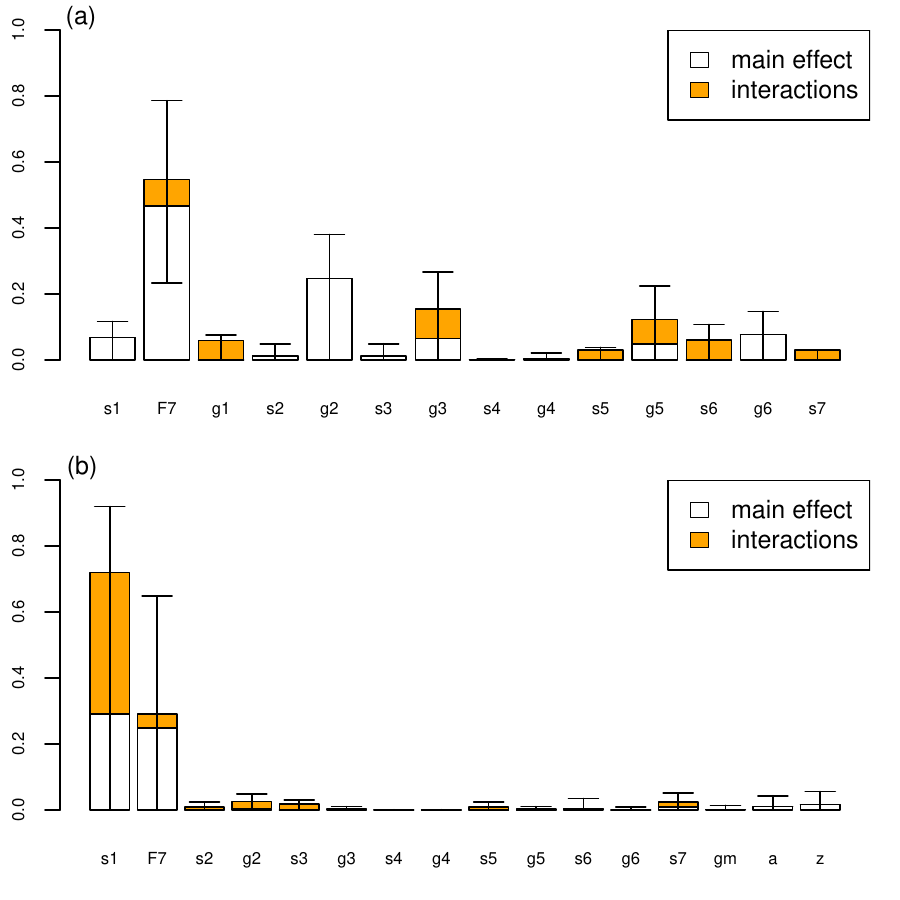}
		\end{center}
		\caption{Sobol indexes for the density independent (a)
		and density dependent (b) models. The bars represent the first and total order
		estimates for the sensitivity of each parameter in the model output. Confidence intervals were generated by bootstrapping 1000 times.}
		\label{fig:sobol}
\end{figure}

The last analyses we present here are the decomposition of variance by Extended Fourier Amplitude Sensitivity Test 
(eFAST, fig. \ref{fig:fast}) and Sobol' methods.
These analyses provides an estimation of the fraction of variation
of model output that can be explained by the individual variation of each parameter 
(which we call first-order sensitivity, or main effect),
along with the total variation caused by interaction between that parameter and others
(total-order sensitivity). The interaction term for each parameter is the difference
between its first and total order sensitivities. 
These analyses are substantially more intensive
in terms of computer time than the previously mentioned. The eFAST analyses presented here required 
7392 and 8448 runs of the simulations, respectively,
for the independent and dependent cases.
Table \ref{tabfast} shows the SBMA measure of concordance between different sizes. Note
that $N_s$ reported should be multiplied by the number of parameters in each model to
obtain the total number of simulations executed. Also note that, while the main effect $D_i$
converge for both model, the total order $D_t$ indexes are still variable with large sample sizes.
This difference in the results for several eFAST analyses, together with 
the differences between the eFAST and Sobol' results, hint at a numerical instability of this method.
Sobol' indexes results are displayed on Fig. \ref{fig:sobol}.

% latex table generated in R 3.2.1 by xtable 1.7-3 package
% Tue Jul  7 00:13:40 2015
\begin{table}[ht]
\centering
\begin{tabular}{rlrrrr}
  \hline
 & Size ($N_s$) & Indep. $D_i$ & Indep. $D_t$ & Dep. $D_i$ & Dep. $D_t$ \\ 
  \hline
1 & 66-132 & 0.78 & 0.85 & -0.14 & 0.52 \\ 
  2 & 132-264 & 0.83 & 0.78 & -0.00 & 0.75 \\ 
  3 & 264-528 & 0.97 & 0.93 & 0.95 & 0.66 \\ 
   \hline
\end{tabular}
\caption{Comparison of eFAST analyses by sample size for both models} 
\label{tabfast}
\end{table}
This analyses reveal that the output of the density independent model is mostly explained
by first-order relations (which explain 62 \% of the 
output variation, according to the eFAST analysis), with $F_7$ and the growth terms being the most important. 
The importance of the linear terms shouldn't come
as a surprise, as the matrix growth model is a linear model.
The density-dependent model, which has 66 \% of 
the output variation predicted by linear terms, according to the eFAST analysis,
exhibit more complex interactions between the terms, but while eFAST indicate the
survival of adults and the terms associated with competition between seedlings as
origins of these interactions, the Sobol' method points at higher-order terms involving
the survival of seedlings.

\subsection{Conclusions}

The results from the uncertainty and sensitivity analyses presented show some of the 
advantages from the methodology described in this work that are unavailable to the 
usual framework used in ecological studies. First, we have been able to quantify the
uncertainty in the asymptotic growth rate (related to $\lambda$) and the stable
population size related to the uncertainty in the model inputs. Also, we provide
a common framework to investigate side-by-side the linear and non-linear matrix models,
from which we were able to point out the similarities and discrepancies between the models.
Analyses based on matrix elasticities are also unable to investigate the role played by
parameters not directly present on the matrix, as the size of the adult trees canopy $z$
in our case.
Finally, our approach allows the identification and quantification of relative importance
of non-linearities and interactions between
the input parameters in determining the model's outcome, and allows us to incorporate
our previous knowledge about the system in specifying the range and distribution 
of each input parameter.

\setkeys{Gin}{width=0.8\textwidth}

\section{Case study 2: Non-linear structured model of {\em Tribolium} population}\label{Tribolium}

\subsection{Model description}
In this section, we present another example of performing uncertainty and sensitivity analyses
to structured population models. This particular example was chosen because of its use by Hal Caswell
in a series of papers published between 2008 and 2010, to illustrate the newly developed 
techniques of analytical sensitivity analyses. Although the theory which provides the tools
to determine sensitivity and elasticity of linear matrix models has been established in the late
1970s, a consistent theory for the study of sensitivity of non-linear models (those in which the
transition frequencies depend on the density or frequency or some size or age class) has only
been fully developed in the recent years \citep{Caswell08, Caswell09, Caswell10}.

In terms of the analytical analyses, the {\em sensitivity} of the result $y$ in respect to a
parameter $x$ is given by the derivative $\frac{dy}{dx}$, and represents the additive effect
that a small perturbation in $x$ exerts over the result $y$. The {\em elasticity} of $y$ in
respect to $x$ is given by $\frac{x}{y}\frac{dy}{dx}$, and represents the proportional effect 
of this perturbation.

To be able to compare the analytical results to a stochastic analysis based on Latin Hypercubes,
we need to define how to estimate the elasticity using this methodology. The formula $\frac{x}{y}\frac{dy}{dx}$
needs to be adjusted in our context for two reasons: first, the derivative must be estimated with some
numerical calculus, and second, the fraction $\frac{x}{y}$ is meaningless as there are no privileged points
to base our analyses on. We propose the following formula as a candidate definition for the
elasticity of $y$ in relation to $x$, evaluated with a stochastic procedure:

\begin{equation}
	\frac{\langle x \rangle}{ \langle y \rangle}s_{yx}
\end{equation}

Here, the brackets $\langle \rangle$ represent the average of a function, and
$s_{yx}$ is the linear partial correlation coefficient of $y$ in relation to $x$.
We stress that this is an arbitrary decision, but as we will show in the following subsection,
it agrees with the analytical results for the investigated model.

The transition matrix for the {\em Tribolium} model is given by:

\begin{equation}
	\mathbf{A}[\mathbf{\theta},\mathbf{n}]  = \left[
		\begin{array} {ccccccc}
			0 &   0 &   b \exp(-c_{el}n_1-c_{ea}n_3) \\
			1 - \mu_l & 0 & 0 \\
			0 & \exp(-c_{pa}n_3) & 1-\mu_a 
		\end{array}
		\right]
		\label{TribLefMatrix}
\end{equation}

Here, $\mathbf{n}(t)$ is the vector representing the beetle population, divided in three life stages:
larvae, pupae and adult. The vector $\mathbf{\theta}$ represents the model parameters, I.e., the vital rates
used in the model.

The non-zero elements of this matrix, from left to right and from top to bottom, are:
\begin{itemize}
	\item Adult fecundity, given by clutch size $b$ times a term of cannibalism of eggs by
		adults (at a rate $c_{ea}$) and by larvae (at a rate $c_{el}$);
	\item Maturation of larvae, reduced by base death rate $\mu_l$;
	\item Pupae eclosion, reduced by cannibalism from adults at a rate $c_{pa}$ 
		(base pupae death rate is effectively zero);
	\item Permanence in the adult class, reduced by base adult mortality rate $\mu_a$.
\end{itemize}

The best estimator for the parameter values, given by the original paper\citep{Dennis1995}, is

\begin{Schunk}
\begin{Sinput}
> b = 6.598
> cea = 1.155e-2
> cel = 1.209e-2
> cpa = 4.7e-3
> mua = 7.729e-3
> mul = 2.055e-1
\end{Sinput}
\end{Schunk}

Also, the original paper focuses on the metabolic equivalent of the beetle population from
different life stages, which is given by $N_m(t) = \mathbf{c}^T\mathbf{n}(t)$, 
with $\mathbf{c}^T = ( 9, 1,  4.5) \mu l CO_2h^{-1}$. Thus, we will focus on $N_m(t)$, 
which is a scalar quantity.

Using the parameters given above, the model converges to a stable fixed point in which
$N_m(t) = 1952$.

\subsection{Elasticity analyses}
\begin{figure}[h!]
\includegraphics{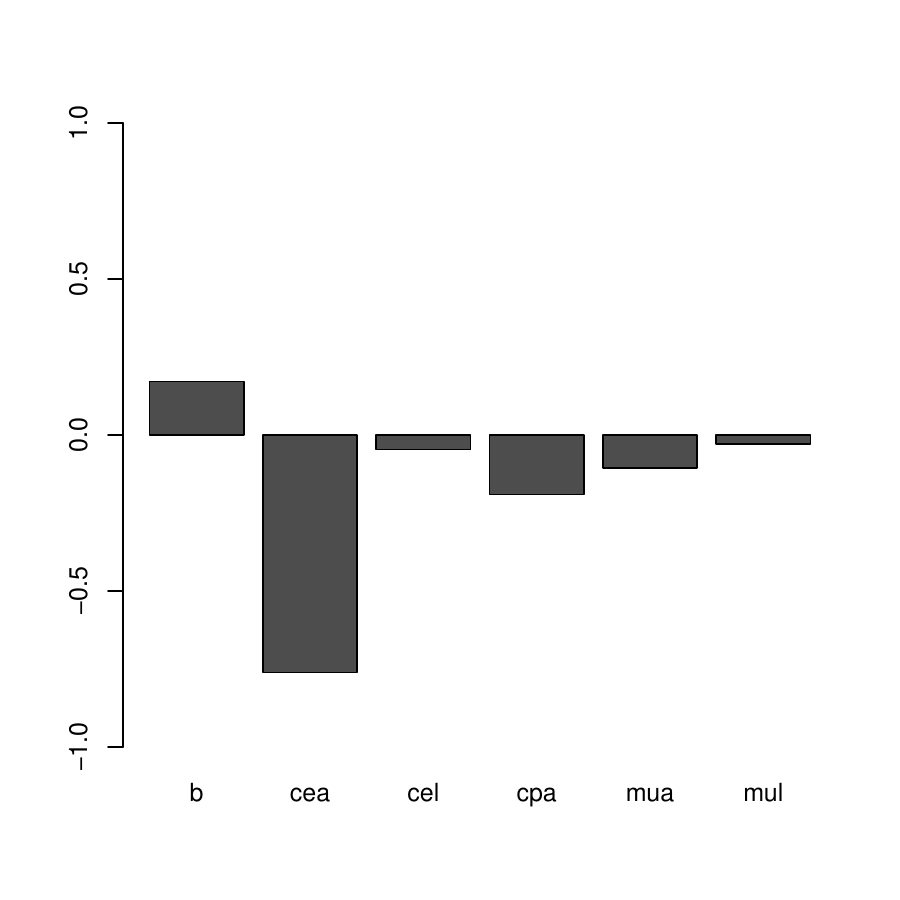}
	\caption{Analytical elasticity analysis for the structured population growth
	model of {\em Tribolium} beetles. Bars represent the elasticity of the metabolic
	equivalent of the equilibrium population in respect to each parameter.}
	\label{analitico}
\end{figure}

The results of analytical elasticity analysis, following \citep{Caswell08} (and replicating
the figure displayed in that paper), are displayed on figure \ref{analitico}\footnote{
The calculations are worked out step by step in the reference cited}.

A raise in the beetle's clutch size causes a positive change on
the final value of $N_m(t)$. All other parameters have negative
elasticities, with $c_{ea}$ having the greatest impact.

\begin{figure}
\includegraphics{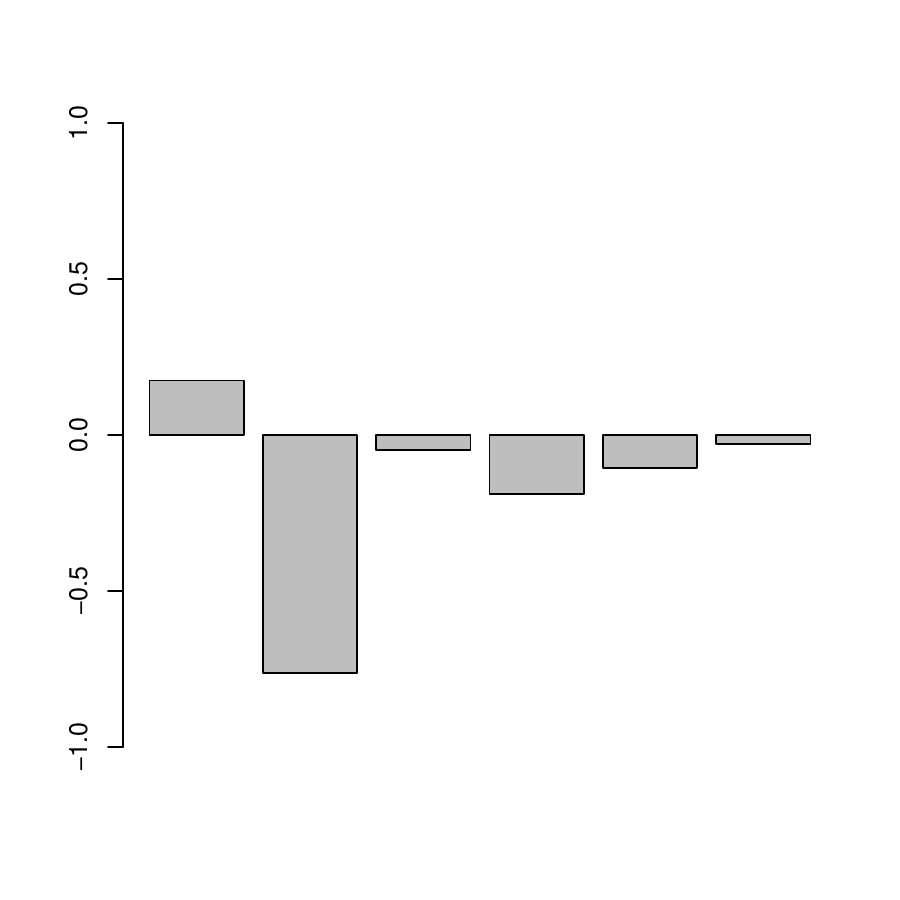}
	\caption{Stochastic elasticity analysis for the structured population growth
	model of {\em Tribolium} beetles, assuming small perturbations. 
	Bars represent the elasticity of the metabolic
	equivalent of the equilibrium population in respect to each parameter.}
	\label{LHSpeq}
\end{figure}

After this analyses, we proceed to a stochastic exploration of the
parameter space with the Latin Hypercube. All parameters are supposed to be
normally distributed with a small dispersion (standard deviation of 1e-08).
The result, presented on figure \ref{LHSpeq}, shows that there is a good correspondence
between both methods.

However, the analytical methods due to Caswell are limited to a very narrow neighborhood
of the estimated parameter vector. If the measurement error is large enough that
the linear approximation of the matrix becomes invalid, other methods are needed in order
to study the sensitivity of the model. A purely analytical possibility is to take into account
higher order derivatives of the matrix terms; however, that leads to a very fast growth of the
complexity of the calculations. The stochastic approach to estimate sensitivity and elasticity 
of the parameters has the advantage of being performed in exactly the same way, regardless of 
how large is the uncertainty of the input parameters. 

\begin{figure}
\includegraphics{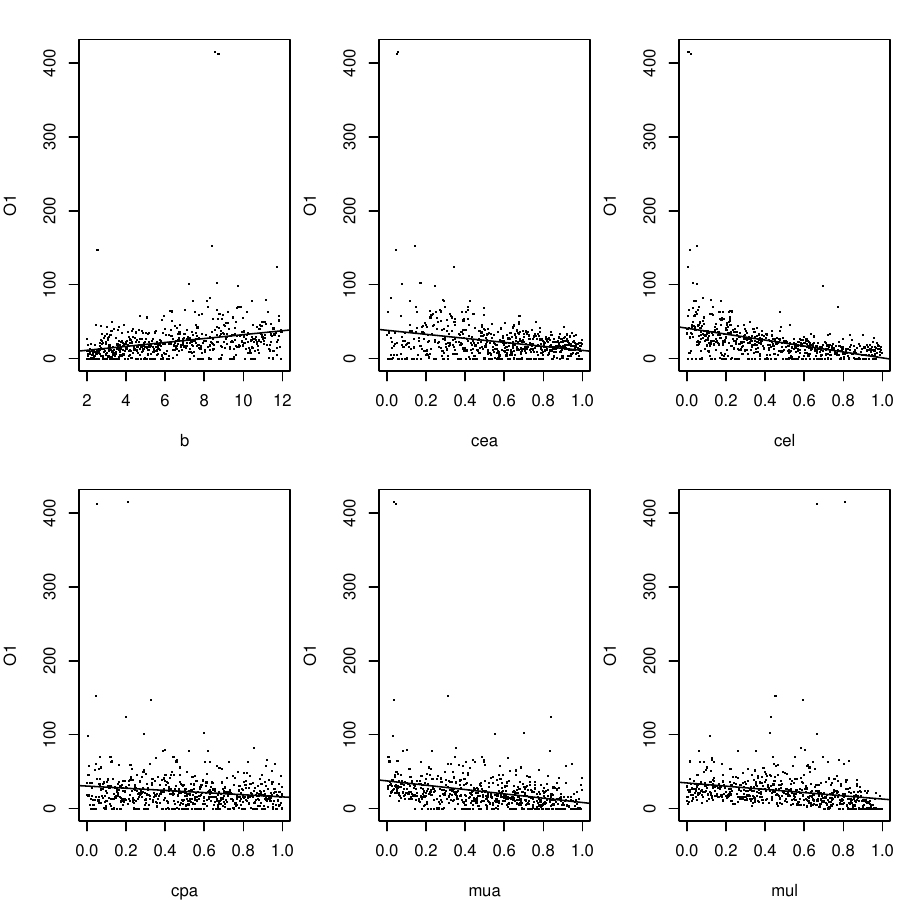}
	\caption{Scatterplots of the metabolic equivalent of the {\em Tribolium}
	beetle as a function of large changes in the input parameters for the population
	growth model}
	\label{corPlot}
\end{figure}

We have repeated the same analysis, but now with an uniform distribution of the parameters
with very large ranges (from $0$ to $1$ in the rates, and from $2$ to $12$ in the clutch 
size), and have found out a much more complex figure, including non-linear and interaction
terms between the parameters. Figure \ref{corPlot} shows the scatter plots, presenting a 
strong nonlinear response to the $c_{ea}$ parameter and possibly complex interactions 
between parameters. In this analysis, we have excluded simulations where the population
did not converge after 2000 time steps.

\begin{figure}
\includegraphics{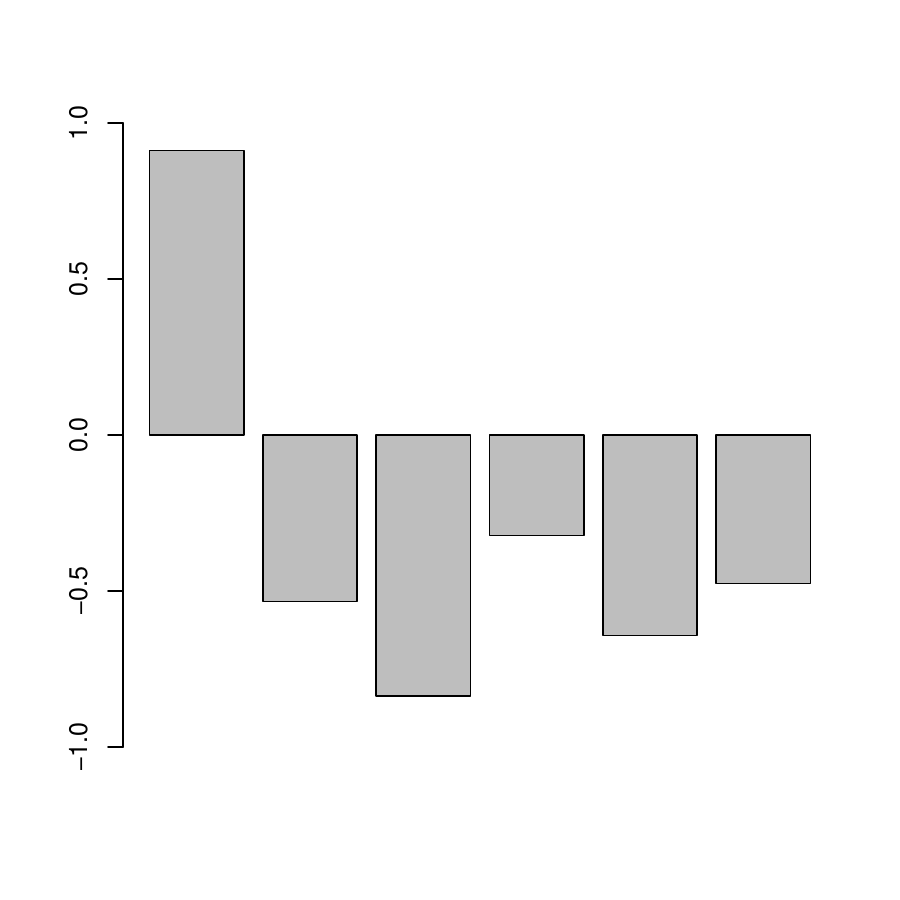}
	\caption{Stochastic elasticity analysis for the structured population growth
	model of {\em Tribolium} beetles, assuming large perturbations. 
	Bars represent the elasticity of the metabolic
	equivalent of the equilibrium population in respect to each parameter.}
	\label{LHSlge}
\end{figure}

The corresponding elasticity analysis, now with a less restricted parameter space,
are shown on figure \ref{LHSlge}. Even if no elasticity has changed its direction,
all present a marked difference between the small and large perturbation scenarios.
The biggest change occurs on parameter $\mu_l$, which raises by 1509\%. 

Given this marked difference between the values, it is necessary to remember that this
contradiction does not mean that one method is right while the other is wrong:
each method is answering a different question.

\newpage
\section{Previous use of Latin Hypercube in ecology}\label{Studies}
\begin{figure}[htpb]
		\begin{center}
				\includegraphics[width=263px]{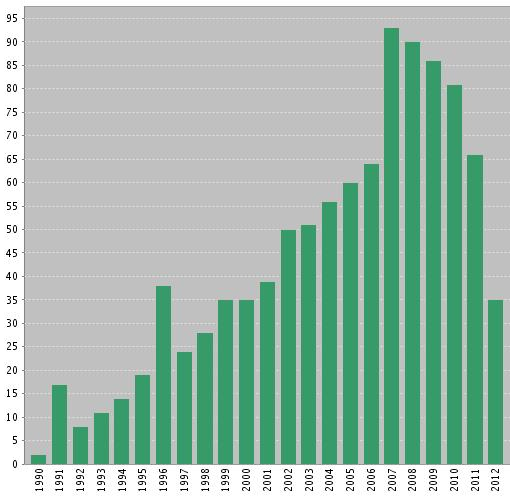}
				\includegraphics[width=243px]{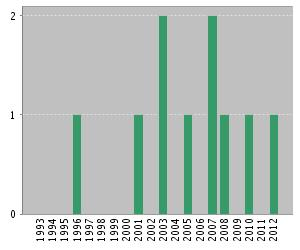}
		\end{center}
		\caption{Top: Number of papers per year since 1990 
		containing the topics ``Sensitivity 
		Analysis, Uncertainty Analysis or Parameter Space Exploration'' 
		in the journals American Naturalist, Ecology, Journal of 
		Ecology, Oikos, Oecologia, Ecological Modelling, 
		Ecology Letters, Journal of Theoretical Biology as reported by
		Thomson Reuter's Web of Knowledge.
		Bot: Restriction of the search above to the keywords ``Latin
		Hypercube''.
		Search conducted 18.06.2012,
		14h GMT.}
		\label{fig:wok}
\end{figure}

The importance of sensitivity and uncertainty analyses in the development 
and use of ecological models is widely recognized. Searching for the terms
``Sensitivity Analysis, Uncertainty Analysis or Parameter Space
Exploration'' in the Web of Knowledge reports 1199 papers in eight major
journals (see fig. \ref{fig:wok} and legend for details) since 1971, 
with 31650 total
citations. However,
most of these papers rely on full and individual parameter space
exploration, which, as discussed in section \ref{Sampling}, are not
optimal. When restricting those results with the keywords ``Latin Hypercube,
MCMC, Markov, Monte Carlo'', just 120 works show up in the results. Of 
those, only 13 (about $10\%$) use Latin Hypercube Sampling 
\citep{Berthaume12, Confalonieri10, Meyer07, Tiemeyer07, Xu05,
Moore04, Shirley03, Duchesne03, Reed84, Marino08, Nathan01, Hamilton10,
Lovvorn96}. There are also relevant examples of LHS use in other
journals \citep{Estill12, Fisher10, Thebault10}.

Also, many of these papers did not explicitly take into account the
correlations between parameters. Those who did used mostly Iman and 
Conover's method \citep{ImanConover82}. 

These works have used Latin Hypercubes typically from 10 to 30 dimensions,
but ranging from 6 to 143 \citep{Berthaume12},
and the number of
simulations ranged from 19 \citep{Nathan01} to 2000 \citep{Tiemeyer07}. 
Also, these works are from varied areas within ecology: applied plant 
ecology \citep{Confalonieri10, Tiemeyer07}, species richness 
\citep{Hamilton10}, epidemiology \citep{Shirley03} and food chain analysis
\citep{Duchesne03}, stressing that the method is useful on varied 
problems.

\section*{Acknowledgements}
We would like to thank Camila Castanho, Charbel El-Hani, Jo\~ao Luis Ferreira Batista, Alexandre Adalardo,
Marina C. Salles and several others for the insightful comments.

This work was supported by a CAPES scholarship.

\newpage
\bibliographystyle{apalike}
\bibliography{chalom}

\end{document}